\documentclass[11pt,superscriptaddress,nofootinbib]{article}
\usepackage{cite}
\usepackage{mathptmx}
\usepackage{amsmath,amsfonts,amssymb}%,calrsfs}
\usepackage[small,bf,hang]{caption}
\usepackage[bookmarks=true,hyperfigures=true,hidelinks]{hyperref}
\usepackage{slashed}
\usepackage{latexsym,epsfig}
\usepackage{color}
\usepackage{mathtools}
\usepackage{nicefrac}
\usepackage{mathrsfs}

\usepackage{lmodern}
\usepackage{simplewick}
\usepackage{xcolor}

\newcommand{\tr}{\mathrm{tr}}

\def\hybrid{
        \topmargin -20pt
        \oddsidemargin 0pt
        \headheight 0pt \headsep 0pt
        \textwidth 6.25in % A4 paper
        \textheight 9.5in % A4 paper
        \marginparwidth .875in
        \parskip 5pt plus 1pt \jot = 1.5ex}

\hybrid

 \csname
@addtoreset\endcsname{equation}{section}

\def\moth{\mathsurround=0pt}
\newdimen\zo \zo=0pt

\def\tick{\leaders\hrule height 0.5ex depth 0pt \hskip 0.5pt}
\def\upboxfill{$\moth \setbox\zo\hbox{\tick}%
  \hskip 3pt\hbox to 0pt{$\tick$\hss}\hrulefill \hbox to 7.5pt{$\tick$\hss}$}

\def\dtick{\leaders\hrule height .34pt depth 0.5ex \hskip 0.5pt}
\def\downboxfill{$\moth \setbox\zo\hbox{\dtick}%
  \hskip 2pt\hbox to 0pt{$\dtick$\hss}\hrulefill \hbox to 2pt{$\dtick$\hss}$}

\def\md{\mathrm{d}}

\def\natural{{\mathbb N}}

\newcommand{\sfrac}[2]{{\textstyle\frac{#1}{#2}}}

\def\bec{\begin{center}}
\def\ec{\end{center}}
\def\a{\alpha}  

\def\b{\beta}  
\def\c{\gamma} 
\def\C{\Gamma}
\def\d{\delta} 
\def\D{\Delta}
 
\def\ve{\varepsilon}

\def\l{\lambda}

\def\m{\mu}
\def\n{\nu}
\def\r{\rho}
\def\s{\sigma}

\def\x{\xi}

\def\pa{{\partial}}

\def\ra{{\rightarrow}}

\def\cD{{\cal D}}

\def\cO{{\cal O}}

\def\cN{{\cal N}}
\def\cR{{\mathcal{R}}}

\def\cT{{\cal T}}

\def\cO{{\cal O}}

\def\ra{\rightarrow}

\def\q{\quad}
\def\qq{\quad\quad}
\def\qqq{\quad\quad\quad}

\def\tr{{\rm Tr}}

 \def\det{{\rm det\,}}
\def\be{\begin{equation}}
\def\ee{\end{equation}}
\def\bea{\begin{eqnarray}}
\def\eea{\end{eqnarray}}
\def\ba{\begin{array}}
\def\ea{\end{array}}

\begin{document}

\begin{titlepage}
\begin{flushright}    
{\small $\,$}
\end{flushright}
\vskip 1cm
\centerline{\LARGE{\bf{ Perturbative linearization of}}}
\vskip 0.5cm
\centerline{\LARGE{\bf{supersymmetric Yang--Mills theory}}}
\vskip 1.5cm
\begin{center}
\begingroup\scshape\Large
Sudarshan Ananth$^{\,\dagger}$, Olaf Lechtenfeld$^{\,\diamond}$, Hannes Malcha$^{\,\star}$,\\[8pt]
Hermann Nicolai$^{\,\star}$, Chetan Pandey$^{\,\dagger}$ and Saurabh Pant$^{\,\dagger}$
\endgroup
\end{center}
\vskip 1.5cm
\centerline{{\Large ${\,}^\dagger$}\it {Indian Institute of Science Education and Research}}
\centerline{\it {Pune 411008, India}}
\vskip 0.5cm
\centerline{{\Large ${\,}^\diamond$}\it {Institut f\"ur Theoretische Physik and Riemann Center for Geometry and Physics}}
\centerline {\it {Leibniz Universit\"at Hannover, Appelstrasse 2, 30167 Hannover, Germany}}
\vskip 0.5cm
\centerline{{\Large ${\,}^\star$}\it {Max-Planck-Institut f\"ur Gravitationsphysik (Albert-Einstein-Institut)}}
\centerline {\it {Am M\"{u}hlenberg 1, 14476 Potsdam, Germany}}
\vskip 1.5cm
\centerline{\bf {Abstract}}
\vskip .5cm
\noindent
Supersymmetric gauge theories are characterized by the existence of a transformation of the bosonic fields 
(Nicolai map) such that the Jacobi determinant of the transformation equals the product of the 
Matthews--Salam--Seiler and Faddeev--Popov determinants. This transformation had been 
worked out to second order in the coupling constant. In this paper, we extend this result (and the 
framework itself) to third order in the coupling constant. A diagrammatic approach in terms of
tree diagrams, aiming to extend this map to arbitrary orders, is outlined. 
This formalism bypasses entirely the use of anti-commuting variables, as well as
issues concerning the (non-)existence of off-shell formulations for these theories. It thus
offers a fresh perspective on supersymmetric gauge theories and, in particular, 
the ubiquitous $\mathcal N=4$ theory.
\vfill
\end{titlepage}

\section{Introduction and summary}

A key consequence of supersymmetry is the dramatic improvement it produces in 
the ultra-violet behavior of quantum field theories. The importance of anti-commuting 
variables in formulating and quantizing supersymmetric field theories is well known. 
These variables are, however, not the easiest to work with, especially 
when it comes to setting up an off-shell formulation,
which often leads to a proliferation of auxiliary variables and unphysical degrees of
freedom (and for the most interesting theories is not believed to exist). The idea that 
supersymmetric theories could be formulated without anti-commuting variables, 
and thus characterized in a more economical fashion in terms of a purely bosonic
functional measure was first proposed in \cite{Nic1,Nic2} and further developed 
in\cite{FL,DL1,DL2,L1}.\footnote{
We emphasize that quantization is an essential and indispensable feature of this formulation. 
An appropriate (perturbative) regularization of all relevant expressions, and in particular of (\ref{eq:MainResult}) below, 
can be obtained by replacing all propagators $C(x)$ by $C_\kappa(x)$ with a cutoff parameter $\kappa$ 
and by  introducing appropriate $\kappa$-dependent multiplicative renormalizations for the coupling constant and the gauge field.}  
This approach to supersymmetric fermion-boson models, referred to as a `Nicolai map',  
is designed to cancel out the fermion determinant while simultaneously reducing the boson measure to a free one 
(see~\cite{JK} for a pedagogical introduction).
Supersymmetric gauge theories, the focus of this paper, can be characterized by the existence of such a functional map $\cT_g$ -- 
a transformation of the bosonic fields such that the Jacobi determinant of the transformation exactly equals the product of 
the Matthews--Salam--Seiler (MSS)~\cite{MS,S} and Faddeev--Popov (FP)~\cite{FP,tH} determinants. This was explicitly 
shown to second order in the coupling constant in \cite{Nic1}. In this paper, we extend these results 
and the framework itself to third order in the coupling constant by presenting an 
explicit formula for $\cT_g$, in equation (\ref{eq:MainResult}). In addition, we provide 
a Feynman-like graphical approach using tree diagrams that in principle allows one to extend this construction to all orders.

\vskip 0.3cm
\noindent This novel approach sidesteps entirely the use of abstractly defined anti-commuting objects and hence offers a fresh perspective 
on the quantization of  supersymmetric gauge theories. Given the central role quantum gauge theories play in describing 
the real world, a new window into their workings is invaluable. This framework, for example, allows us to re-derive~\cite{ANPP} 
the classical result~\cite{SYM1} that interacting supersymmetric Yang--Mills theories exist only in space-time dimensions $D=3,4,6$ and $10$ 
(together with the extended supersymmetric Yang--Mills theories obtained from these by dimensional reduction). 
Specifically, for the $\mathcal N=4$ theory in $D=4$, all known results for scalar correlators may also be obtained within this formalism, 
at least up to the order for which the map had previously been worked out \cite{NP}; the computational efforts involved compare well 
with the more standard techniques of perturbative quantum field theory. An understanding of scattering amplitudes in this approach is likely 
to yield new insights into the symmetries that underlie these simple structures. A related long-term goal is to move beyond perturbation theory 
and establish a link between the map and the integrable properties of the $\mathcal N=4$ model (see {\em e.g.} \cite{Beisert}). 
There also exists earlier work which focussed primarily on the search for, and the exploitation of, variables providing a {\em local\/} realization 
of the map $\cT_g$ \cite{Fub1,Fub2, Fub3,FLMR,AFF,Boc1} (the precise relation between these older results and the {\em non-local\/} map $\cT_g$ 
will be left for future study), as well as an alternative construction of $\cT_g$ in \cite{L2}.
\vskip 0.3cm
\noindent 
We start by stating the main theorem from~\cite{Nic1,Nic2}.

\subsection{Main theorem}
\label{th:MainTheorem}
{\it {Supersymmetric gauge theories are characterized by the existence of 
a non-linear and non-local transformation $\cT_g$ of the Yang--Mills fields
\begin{align*}
\cT_g \, : \, A_\m^a(x)\ \mapsto\ A_\m^{\prime \, a}(x,g;A) \, , 
\end{align*}
which is invertible at least in the sense of a formal power series such that
\begin{enumerate}
\item 
The bosonic Yang--Mills action without gauge-fixing terms is mapped to the abelian action,
\begin{align*}
S_0[A']\ =\ S_g[A] \, ,
\end{align*}
where $S_g[A] \equiv \frac14\int \mathrm{d}x  \, F_{\m\n}^a F_{\m\n}^a$ with non-abelian 
field strength  $F_{\m\n}^a \equiv \pa_\m A_\n^a - \pa_\n A_\m^a + gf^{abc} A_\m^b A_\n^c$
and with gauge coupling $g$; $S_0$ denotes the free action ($g=0$). 
\item 
The gauge-fixing function $G^a(A)$ is a fixed point of $\cT_g$.
\item  
Modulo terms proportional to the gauge-fixing function $G^a(A)$,
the Jacobi determinant of $\cT_g$ is equal to the product of the MSS and FPP determinants~\footnote{
With the understanding that $\D_{MSS}$ is really a Pfaffian for Majorana fermions.} 
\begin{align*}
\det \left( \frac{\d A_\m^{\prime \, a}(x,g;A) }{\d A_\n^b(y)} \right)\ =\ \D_{MSS} [A]\ \D_{FP}[A] \, ,
\end{align*}
at least order by order in perturbation theory.
\end{enumerate}}}
\noindent 
A general proof of this theorem is presented in appendix A, and is largely based on existing work~\cite{Nic2,FL,DL1,DL2,L1}. 
However, the present proof is more general in that it takes into account recent insights from~\cite{ANPP} 
that the theorem is actually valid for {\em all\/} supersymmetric Yang--Mills theories in space-time dimensions $D=3,4,6,10$. 
These include the corresponding extended theories obtained by reduction (like $\mathcal N=4$ Yang--Mills theory from the $D=10$ theory). 
Although the above theorem applies with arbitrary gauge groups
we will for simplicity in the remainder restrict our attention to SU($n$), with real antisymmetric
structure constants $f^{abc}$ obeying
\begin{equation}\label{ff}
f^{abc} f^{abd} = n\,\delta^{cd} \, .
\end{equation}
%\vskip 0.3cm

\noindent 
Even with the general proof and explicit expressions for $A_\mu^{\prime \, a}(x)$ at hand it is by no means obvious 
that the transformed field $A_\mu^{\prime \, a}(x)$ satisfies all three statements 
in the main theorem above, and at every order in $g$.  
Our main goal with this paper is to work out the transformation at $\cO(g^3)$ and to explain in detail 
how all the necessary conditions are satisfied. We hope that the explicit expressions, derived in 
section \ref{sec:Tests}, illustrate the non-triviality of this result. With these findings 
one can now proceed to compute quantum correlators:
we have
\begin{equation}
\Big\langle\!\!\!\Big\langle A_{\mu_1}^{a_1} (x_1) \cdots A_{\mu_n}^{a_n}(x_n) \Big\rangle\!\!\!\Big\rangle_g \ = \, 
\Big\langle A_{\mu_1}^{a_1} (x_1) \cdots A_{\mu_n}^{a_n}(x_n) \Big\rangle_g \, ,
\end{equation}
where for any monomial $X[A]$ \footnote{We note that in principle the formalism can 
also be extended to fermionic correlators by admitting non-local expressions for $X[A]$.
}
\begin{equation}\label{corr0}
\begin{aligned}
\big\langle \!\!\big\langle X[A] \big\rangle\!\!\big\rangle_g 
&\ := \ \int \cD A\,\cD\lambda\,\cD C\,\cD \bar{C} \
\mathrm{e}^{- S_\text{inv}[A,\lambda]-S_\text{gf}[A,C,\bar{C}] } \ X[A] \, ,\\[2mm] 
\big\langle X[A] \big\rangle_g &\ := \ \int \cD_g[A] \ X[A] \, ,
\end{aligned}
\end{equation}
and $S_\text{inv}[A,\lambda]$ and $S_\text{gf}[A,C,\bar{C}]$ denote the supersymmetric
Yang--Mills action and the gauge-fixing action defined in (\ref{Sinv}) and (\ref{Sgf}),
respectively. Furthermore, $\cD_g[A]$ denotes the (non-local) bosonic  
functional measure of the interacting  theory obtained  after integrating out 
all anti-commuting variables (the gauginos $\lambda^a(x)$ and the ghosts
$C^a(x)$ and $ \bar{C}^a(x)$).  Observe that all these expectation values are already 
properly normalized (that is, 
$\langle\!\!\langle 1 \rangle\!\!\rangle_g = \langle 1 \rangle_g = 1$ for all $g$) 
by the vanishing of the vacuum energy in supersymmetric theories (this statement
remains true with the properly normalized gauge-fixing action $S_\text{gf}$).
The theorem then tells us that, again modulo terms proportional to the
gauge-fixing function,
\begin{equation}\label{corr1}
\Big\langle A_{\mu_1}^{a_1} (x_1) \ \cdots\ A_{\mu_n}^{a_n}(x_n) \Big\rangle_g
 \ =\ \Big\langle (\cT^{-1}_gA)_{\mu_1}^{a_1}(x_1)\ \cdots \
 (\cT_g^{-1}A)_{\mu_n}^{a_n}(x_n) \Big\rangle_0
\end{equation}
%\vskip 0.3cm

\noindent 
Let us also emphasize that, at this point, all the statements in the above theorem are to be understood 
in the sense of formal power series. Non-perturbatively, we will have to worry 
about zero modes of the relevant determinants \cite{Nic82}: on the mass shell (where
$\delta S_g /\delta A = 0$) we have
\begin{equation}\label{d2S}
\frac{\d^2 S_g[A]}{\d A_\mu^a(x)\,\d A_\nu^b(y)} \ =\
\int \mathrm{d}w \ \mathrm{d}z \ \frac{\d A^{\prime \, c}_\a(w)}{\d A_\mu^a(x)}
\big(- \Box\, \d_{\a\b} + \pa_\a \pa_\b \big) \d(w-z) \frac{\d A^{\prime\,  c}_\b(z)}{\d A_\nu^b(y)}\, ,
\end{equation}
directly relating the Jacobian of the transformation to the second-order fluctuation 
operator around a given background solution of the Yang--Mills field equations, 
such as an instanton solution (the addition of the second variation of the gauge-fixing
action $S_\text{gf}$ renders the integration kernel on the r.h.s. of (\ref{d2S}) invertible).
Since the eigenvalues of the fluctuation operator are known to be related 
to the ones of the MSS determinant, matching zero modes imposes an extra 
restriction on the theory. As already pointed out in \cite{Nic82} 
the match works only for the maximally extended $\mathcal N=4$ theory in $D=4$, which is additionally 
singled out for this reason.

\vskip 0.3cm
\noindent 
In the remainder we restrict ourselves to the Landau (or classically: Lorenz) gauge, that is,
\begin{align}\label{Landau}
G^a(A) \ = \ \pa^\m A_\m^a \, .
\end{align}
The second statement of the theorem then means that the longitudinal part of the
gauge field is not affected by the map $\cT_g$, and therefore is the same as in the free theory.
Finally, other choices for the gauge function $G^a(A)$,  in particular the axial and light-cone 
gauges, are possible, but will be discussed elsewhere. The light-cone gauge is of
particular interest, not only because it is the only gauge in which the UV finiteness of 
the maximal $\cN=4$ theory is manifest \cite{M,BLN}, but also because for pure and maximally 
supersymmetric Yang--Mills theories the associated Hamiltonians are quadratic forms~\cite{ABM,APP}. 
This ``complete square" structure, highly reminiscent of the map, 
is likely related to a light-cone realization of~$\cT_g$.

\subsection{Conventions and notations}
The work presented below is in Euclidean space, rendering upper and lower 
indices equivalent. However, the Euclidean metric is by no means crucial to our 
discussion, as all of these results can be derived in spacetimes with a Lorentzian 
signature as in~\cite{FL, DL1, DL2}. For the $\gamma$-matrices we thus have 
the usual Clifford algebra relation $\{\gamma_\mu , \gamma_\nu\} = 2\delta_{\mu\nu}$.
In order not to over-clutter the notation, we will not explicitly distinguish between
Majorana, Weyl, and Majorana-Weyl spinors. This notational shortcut is justified because
for our calculations all we need is the basic Clifford algebra relation and the trace
(in the sense of a formal algebraic prescription)
\begin{equation}\label{id}
{\rm Tr}\,\, {\bf{1}} \,=\, r
\end{equation}
where $r$ counts the number of off-shell fermionic degrees of freedom for the cases 
of interest. The extra factor of $1/2$ required for the on-shell degrees of freedom is included
in our perturbative definition of the MSS determinant in (\ref{eq:MSSDet}).\footnote{
Because we use only the Clifford algebra and the normalization (\ref{id})
we also need not worry about issues with the existence or non-existence of Euclidean Majorana 
spinors. Following \cite{Nic78}, the Euclidean supersymmetric theory here {\em by definition\/} is
the one whose correlators coincide with the analytic continuation of Lorentzian correlators
to imaginary time, in accordance with the Osterwalder-Schrader reconstruction theorem \cite{OS}.
}
With the admissible space-time dimensions $D=3,4,6$ and $10$ and the representation dimensions $r=2,4,8$ and $16$ of the corresponding Clifford algebras, 
we then have the relation
\begin{align}\label{rD}
r = 2(D-2) \, .
\end{align}
In section \ref{sec:Tests} and the appendix~A.2 we will rederive this relation in a novel manner.

\vskip 0.3cm
\noindent
We shall employ the generator~$\cR$ of the inverse transformation $\cT^{-1}_g$ (see below).
A key  role in the $\cR$ operator is played by the fermionic propagator $S^{ab}(x,y;A)$ 
in a gauge-field dependent background characterized by $A_\m^a(x)$, with
\begin{align}
\c^\m \left( D_\m S \right)^{ab}(x,y;A)\ \equiv \
\c^\m \left[ \d^{ac} \pa_\m\,+\,g f^{adc} A_\m^d(x) \right] S^{cb}(x,y;A)\ =\  \d^{ab} \d(x{-}y) \, .
\label{eq:DefFermProp}
\end{align}
The limit $g = 0$ gives us the free fermionic propagator $S_0^{ab}(x)$ which obeys
\begin{align}
\c^\m \pa_\m S_0^{ab}(x) \ =\ \d^{ab} \d(x) \qq \Rightarrow \qq S_0^{ab}(x)\  =\  - \d^{ab} \c^\m \pa_\m C(x) \, .
\end{align}
Here $C(x)$ is the free scalar propagator (with the Laplacian $\pa_\m \pa^\m \equiv \Box$)
\begin{align}
C(x) \ =\ \int \frac{\md k}{(2\pi)^D} \frac{\mathrm{e}^{\mathrm{i}kx}}{k^2} 
\ =\ \frac{1}{(D{-}2)\,D\,\pi^{D/2}} \, \C\bigl( \sfrac{D}{2} + 1 \bigr) (x^2)^{1-\frac{D}{2}}\, .
\end{align}
It satisfies $\Box \, C(x) = - \d(x)$. 
We use the convention that the derivative always acts on the first argument, 
{\em i.e.}~$\pa_\r C(x{-}y) \equiv (\pa / \pa x^\r ) C(x{-}y) \equiv \pa_\r^xC(x{-}y)$. 
Thus, we need to be careful with sign flips when using $\pa_\r^x C(x{-}y) = - \pa_\r^y C(x{-}y) = \pa_\r^x C(y{-}x) = - \pa_\r^y C(y{-}x)$. 

\vskip 0.3cm
\noindent
Finally, we require the implementation of the ghost propagator $G^{ab}(x,y;A)$, obeying
\begin{align}
\pa^\m (D_\m G)^{ab}(x,y;A) \ \equiv\
\left[ \d^{ac} \Box + g f^{adc} 
\frac{\partial}{\pa x^\mu} A_\m^d(x) \right] G^{cb}(x,y;A)\ =\ \d^{ab} \d(x{-}y) 
\label{eq:DelGhostProp}
\end{align}
where the differential operator acts on everything to its right.
Hence, the free ghost propagator satisfies
\begin{align}
\Box G_0^{ab}(x)\ =\ \d^{ab} \d(x) \qq \Rightarrow \qq G_0^{ab}(x) \ =\ -\d^{ab} C(x)\, ,
\end{align}
and the full ghost propagator expands as
\begin{equation}
G^{ab}(x,y) \ =\  G_0^{ab}(x,y) \,-\,  g \int \md z \, G_0^{ac}(x,z) f^{cde} \pa^\mu_z 
                      \big(A_\mu^d(z)  G_0^{eb}(z,y) \big) \,+\ \cdots \, .
\end{equation}
It is important to note that not only $G^{ab}(x,y;A)$ depends on $g$ and the background field 
$A_\m^a(x)$ but that $(D_\m G)^{ab}(x,y;A)$ does so as well, {\em viz.}~\footnote{ 
We note that the corresponding formula (1.18) in \cite{ANPP} is incomplete in that
it missed out on the $g$-dependence of $(D_\mu G)^{ab}$. However,  
this correction kicks in only at ${\cal{O}}(g^3)$ and beyond, hence
does not affect the results up to second order in \cite{Nic1,ANPP}.} 
\begin{align}
-(D_\m G)^{ab}(x,y;A)\ =\ \d^{ab} \pa_\mu C(x{-}y)\,+\,g f^{acb} \int \md z \ \Pi_{\m \n}(x{-}z) A_\n^c(z) C(z{-}y) 
\ +\ \cO(g^2)  \, , 
\label{eq:GhostProp}
\end{align}
with the abelian transversal projector
\begin{align} \label{abelianprojector}
\Pi_{\m \n}(x{-}z)\ \equiv\ \left( \d_{\m \n} - \frac{\pa_\m \pa_\n}{\Box} \right) \d(x-z)
\  \simeq\ \d_{\m \n} \d(x{-}z) + \pa_\m C(x{-}z)\,\pa_\n^z \, , 
\end{align}
where $\simeq$ means equality in the sense of a distribution. We will later see that the terms of $\cO(g)$ in \eqref{eq:GhostProp} become relevant for the map $\cT_g$ from order $g^3$ onwards.

\subsection{The \texorpdfstring{$\cR$}{R} operator}
To determine the map $\cT_g$ one first constructs its {\em inverse\/} $\cT_g^{-1}$ via
its infinitesimal generator $\cR$, a non-local and non-linear functional
differential operator first introduced for the $\cN=1,D=4$  theory in \cite{FL,DL1,DL2}. 
In general, the $\cR$ operator works for any choice of gauge, but we will here
derive the relevant expressions only for the Landau gauge (\ref{Landau}).
The image $(\cT_gA)_\m^a(x)$ of the map is then obtained order by order in $g$ by formally 
inverting the power series
\begin{align}
(\cT_g^{-1}A)_\m^a(x) \ =\ 
\sum_{n=0}^\infty \frac{g^n}{n!} \, \bigl(\mathcal{R}^n A\bigr)_\m^a(x) \Big\vert_{g=0} \, ,
\label{eq:TInverse}
\end{align}
where $\cR$ is the infinitesimal generator of the {\em inverse\/} map.
Details of the construction of the  $\mathcal{R}$ operator are provided in appendix \ref{sec:FlumeLechtenfeldProof}.1, thus generalizing the original proof 
given in \cite{FL} for $D=4$.   
We use the background-field dependent propagators
defined in (\ref{eq:DefFermProp}) and (\ref{eq:DelGhostProp}). 
In appendix A we will also use the notation
\begin{align}
\bcontraction{}{\l}{^a(x)\ }{\bar{\l}} \l^a(x)\ \bar{\l}^b(y) \ \equiv\
 S^{ab}(x,y;A) \qq \text{and} \qq 
 \bcontraction{}{C}{^a(x)\ }{\bar{C}} C^a(x)\ \bar{C}^b(y) \ \equiv\  G^{ab}(x,y;A)
\label{eq:ContractionPropagators}
\end{align}
to rewrite \eqref{eq:ROperatorVersion1}. Here $\lambda^a(x)$ are the gaugino fields (prior to their
elimination via the MSS determinant), and $C^a(x)$ and $\bar{C}^a(x)$ denote the ghost and anti-ghost fields.\footnote{
Not to be confused with the propagator $C(x)$, which carries no indices.}
For the Landau gauge function (\ref{Landau}) the $\mathcal{R}$ operator is then 
represented by the functional differential operator
\begin{align}
\begin{aligned}
&\mathcal{R} \,=\,  \frac{\md}{\md g} - \frac{1}{2r} \int \md x\,\md y \
\tr \left( \c_\m S^{ab}(x,y;A)\,\c^{\r \l} \right)\,f^{bcd} A_\r^c(y) A_\l^d(y)\,\frac{\d}{\d A_\m^a(x)} \\
&\qq - \frac{1}{2r} \int \md x\, \md z\, \md y \ (D_\m G)^{ae}(x,z;A)\,\pa_\n\,
\tr \left( \c^\n S^{eb}(z,y;A)\,\c^{\r \l} \right)\,f^{bcd} A_\r^c(y) A_\l^d(y) \,\frac{\d}{\d A_\m^a(x)}  \\[4pt]
&\q =\,  \frac{\md}{\md g} - \frac{1}{2r} \int \md x\,\md z\,\md y \  P_{\m\n}^{ae}(x,z)\,
\tr \left( \c^\n S^{eb}(z,y;A)\,\c^{\r \l} \right)\,f^{bcd} A_\r^c(y) A_\l^d(y) \,\frac{\d}{\d A_\m^a(x)} \, ,
\label{eq:ROperatorVersion2}
\end{aligned}
\end{align}
Notice that the first part of the $\cR$ operator (first line on the r.h.s.
of (\ref{eq:ROperatorVersion2})) is {\em gauge independent}, whereas the second line does 
depend on the choice of the gauge-fixing function via the ghost propagator
(see appendix A for an explanation of the origin of this term).
Furthermore, we have introduced the `covariant' (or `non-abelian') transversal projector
\begin{align}
\begin{aligned}
P_{\m\n}^{ab}(x,z;A)&\ =\ \d^{ab}\d_{\m\n}\d(x{-}z) - (D_\m G)^{ab}(x,z;A)\,\pa_\n \\[4pt]
&\ =\  \int \md w\ \Pi_{\m\s}(x{-}w)\,\left[ 
\d^{ab}\d_{\s\n}\d(w{-}z) + g f^{acb}A_\s^c(w)\,C(w{-}z)\,\pa_\n +  \cO(g^2) \right]
\end{aligned}
\label{eq:covproj}
\end{align}
obeying $P* P = P$ and $\partial^\mu P_{\mu\nu}^{ab} = 0$. 
This definition (which, for all we know, has not appeared in the literature) differs from
the abelian one~\eqref{abelianprojector} in that there appears a gauge-covariant derivative and a non-linear dependence
on $A$ on the r.h.s.\ of (\ref{eq:covproj}). It allows for a non-standard (non-linear) separation 
between transversal and longitudinal degrees of freedom, with
\begin{equation}
A_\mu^{\perp a}(x) := \int \md y \, P_{\mu\nu}^{ab}(x,y;A)\,A_\nu^b(y) \qquad\textrm{and}\qquad
A_\mu^{|\! | a}(x) := \int \md y \, (D_\mu G)^{ab}(x,y;A)\,\pa^\nu\!A_\nu^b(y) \, ,
\end{equation} 
such that the more standard abelian (linear) split of $A_\mu^a(x)$ into transversal 
and longitudinal parts is recovered by setting $g=0$. 
Equation~(\ref{eq:ROperatorVersion2}) means that the $\cR$ operation acts only on 
the `covariantly transversal' part of its argument. Consequently, the map $\cT_g$ and its inverse
$\cT_g^{-1}$ affect only the transverse degrees of freedom of the gauge field, whereas
they do not change its longitudinal component, which is
therefore effectively the same as in the free theory.

\vskip 0.3cm
\noindent 
Since the bosonic background field $A_\m^a(x)$ does not depend on $g$, the first application 
of $\mathcal{R}$ to $A_\m^a(x)$ is straightforward. For all higher orders we also need
\begin{align}
&\frac{\md S^{ab}(x,y)}{\md g} \ =\  - \int \md z \ S^{ac}(x,z)\,f^{cmd} A_\m^m(z)\, \c^\m S^{db}(z,y) 
\label{eq:SgDerivative}
\end{align}
and
\begin{align}
\frac{\d S^{ab}(z,y)}{\d A_\m^m(x)} \ =\ - g\,S^{ac}(z,x)\,f^{cmd} \c^\m S^{db}(x,y)
\label{eq:SVariation}
\end{align}
as well as
\begin{align}
&\frac{\md G^{ab}(x,y)}{\md g} \ =\  \int \md z \ G^{ac}(x,z)\,f^{cmd} \overleftarrow{\pa}{\!\!}_z^\mu\, A_\m^m(z)\,G^{db}(z,y)
\label{eq:GgDerivative}
\end{align}
and
\begin{align}
\frac{\d G^{ab}(z,y)}{\d A_\m^m(x)} \ =\ g\,G^{ac}(z,x)\,f^{cmd} \overleftarrow{\pa}{\!\!}_x^\m \,G^{db}(x,y)\, .
\label{eq:GVariation}
\end{align}
These equations are obtained from \eqref{eq:DefFermProp} and \eqref{eq:DelGhostProp}. 
After iteratively computing $\mathcal{R}^n$ for any desired $n$, we set $g=0$, which in particular maps $S^{ab}(x,y)$ to the free propagator $S_0^{ab}(x{-}y)$ and $G^{ab}(x,y)$ to the 
free propagator $G_0^{ab}(x{-}y)$, and finally obtain $(\cT_g^{-1}A)_\m^a(x)$ at $\cO(g^n)$. We shall present $(\cT_g^{-1}A)_\m^a(x)\big\vert_{\cO(g^3)}$ in appendix \ref{sec:R3}.
\vskip 0.3cm
\noindent 
The actual map $\cT_g$  is then obtained by power series inversion. Let
\begin{align}
\cT_g  A  \ =\ \sum_{n=0}^\infty \frac{g^n}{n!} \, T_n A \, .
\end{align}
Expanding $\cT_g^{-1} \cT_g  = {\bf{1}}$ in powers of $g$ and matching 
coefficients we readily obtain
\begin{align}
\begin{aligned}
T_0 A \ &=\ A \,  , \\
T_1 A \ &=\ - \cR\,T_0 A\big\vert_{g=0} \, ,  \\
T_2 A \ &=\  - \cR^2  T_0 A\big\vert_{g=0} - 2 \cR\,T_1 A\big\vert_{g=0} \, , \\
T_3 A \ &=\ - \cR^3 T_0 A\big\vert_{g=0} - 3 \cR^2 T_1 A\big\vert_{g=0}  - 3 \cR\,T_2 A\big\vert_{g=0} \, .
\end{aligned}
\label{eq:inversion}
\end{align}
The explicit expression for $(\cT_gA)_\m^a(x)$ up to $\cO(g^3)$ is provided in the following section.

\newpage
\section{Result and discussion} \label{sec:ResultandDiscussion}
We now present the main new result which is the explicit formula for $\cT_g$ to cubic 
order $\cO(g^3)$~\footnote{As usual, all anti-symmetrizations are with strength one,
such that {\em e.g.} $[ab] = \frac12(ab -ba)$.}
\begin{align}
\begin{aligned} 
{}\!\!\!\!
(\cT_gA)_\m^a(x)\ &=\ A_\m^a(x) \ +\ g\,f^{abc} \int \md y \ \pa_\r C(x-y) A_\m^b(y) A_\r^c(y) \\[2.5mm]
&\q + \, \frac{3g^2}{2} f^{abc} f^{bde} \int \md y \  \md z \ \pa_\r C(x-y) A_\l^c(y) \pa_{[\r} C(y-z) A_\m^d(z) A_{\l]}^e(z) 
\\[2.5mm]
&\q + \frac{g^3}{2} f^{a b c} f^{b d e} f^{c m n} \int \md y \ \md z \ \md w \ \pa_\r C(x-y)  \\
&\qq \times  \pa_\l C(y-z)  A_\l^d(z) A_\s^e(z) \pa_{[\r} C(y-w)  A_\m^m(w) A_{\s]}^n(w)   \\
&\q +  \, g^3 f^{a b c} f^{b d e} f^{d m n}  \int \md y \ \md z \ \md w \ \pa_\r C(x-y) A_\l^c(y) \ \bigg\{  \\
&\hspace{2cm}  + 2 \, \pa_{[\r} C(y-z)A_{\s]}^e(z)   \pa_{[\l} C(z-w)  A_\m^m(w) A_{\s]}^n(w)  \\
&\hspace{2cm} - 2 \,  \pa_{[\l} C(y-z)A_{\s]}^e(z)   \pa_{[\r} C(z-w)  A_\m^m(w) A_{\s]}^n(w) \\
&\hspace{2cm} - \pa_{\s} C(y-z)A_{\s}^e(z)   \pa_{[\r} C(z-w)  A_\m^m(w) A_{\l]}^n(w)  \\
&\hspace{2cm} - 2 \, \pa_{[\s} C(y-z)A_{\m]}^e(z)   \pa_{[\r} C(z-w)  A_\l^m(w) A_{\s]}^n(w) \\
&\hspace{2cm} +  \pa_{[\r} C(y-z) A_\m^e(z)  \pa_{ | \s |} C(z-w)  A_{\l ]}^m(w) A_\s^n(w) \bigg\} \\
&\q + \, \frac{g^3}{3} f^{abc} f^{bde} f^{dmn} \int \md y \ \md z \ \md w \ \bigg\{ \\
&\hspace{2cm} +  2 \,  \pa_\r C(x-y) A_{[\r}^c(y) \pa_{\m]} C(y-z) A_\l^e(z) \pa_\s C(z-w) A_\l^m(w) A_\s^n(w) \\
&\hspace{2cm}  - \pa_\m C(x-y) \, \pa_\r \left( A_\r^c(y)  C(y-z) \right) A_\l^e(z) \pa_\s C(z-w) A_\l^m(w) A_\s^n(w) \bigg\} \\
&\q - \, \frac{g^3}{3} f^{abc} f^{bde} f^{dmn} \int \md y \ \md z   
   \  A_\m^c(x) C(x-y) A_\r^e(y) \pa_\l C(y-z) A_\r^m(z) A_\l^n(z)  \\[2.5mm]
&\q +\;  \cO(g^4)    \, .
\label{eq:MainResult}
\end{aligned}
\end{align}
The first two lines above correspond to the result obtained in \cite{Nic1} and extended to dimensions $D\neq 4$ in \cite{ANPP}. The last two lines are the new terms 
arising from the $g$-dependence of $(D_\mu G)^{ab}$ in (\ref{eq:DelGhostProp}); they are crucial for the fulfillment of the conditions in the main theorem.
While the result up to $\cO(g^2)$ was originally obtained by trial and error in \cite{Nic1}, this becomes tricky at higher orders because the number of terms 
is significantly larger at $\cO(g^3)$ than below. In addition, from the last term we see that new 
structures appear. In the following section we will verify that this result indeed satisfies all 
three statements of the main theorem (subsection \ref{th:MainTheorem}) simultaneously, 
providing a highly non-trivial test.

\section{Tests} \label{sec:Tests}
A general proof that the statements in the main theorem in subsection 
\ref{th:MainTheorem} are true at any order of $g$ is given in appendix~A. 
Since a detailed explanation of how this works up to $\cO(g^2)$ can be found 
in \cite{Nic1,ANPP} we will skip the details of these lower order calculations here, 
and only present the details relevant to the third order in~$g$.

\subsection{Gauge condition}
We first verify that $\pa_\m A_\m^{\prime \, a}(x)  = \pa_\m A_\m^a(x) + \cO(g^4)$. 
Applying $\pa_\m$ to the terms of order $g^3$ in \eqref{eq:MainResult} and removing all 
terms that are manifestly anti-symmetric under the exchange of indices $\mu$ and $\rho$ yields
\begin{align}
\begin{aligned}
\pa_\m A_\m^{\prime \, a}(x) \big\vert_{\cO(g^3)} 
&=   g^3 f^{a b c} f^{b d e} f^{d m n}  \int \md y \ \md z \ \md w \ \pa_\m \pa_\r C(x-y) A_\l^c(y) \ \bigg\{  \\
&\hspace{2cm}+ 2 \, \pa_{[\r} C(y-z)A_{\s]}^e(z)   \pa_{[\l} C(z-w)  A_\m^m(w) A_{\s]}^n(w)  \\
&\hspace{2cm} - 2 \, \pa_{[\s} C(y-z)A_{\m]}^e(z)   \pa_{[\r} C(z-w)  A_\l^m(w) A_{\s]}^n(w)  \bigg\} \\
&\q - \frac{g^3}{3} f^{abc} f^{bde} f^{dmn} \int \md y \ \md z \ \md w  \\
&\qq \times  \Box \,  C(x-y) \, \pa_\r \left( A_\r^c(y)  C(y-z) \right)   A_\l^e(z) \pa_\s C(z-w) A_\l^m(w) A_\s^n(w) \\
&\q - \frac{g^3}{3} f^{abc} f^{bde} f^{dmn} \int \md y \ \md z   \\
&\qq  \times  \pa_\m \left( A_\m^c(x) C(x-y) \right) A_\r^e(y) \pa_\l C(y-z) A_\r^m(z) A_\l^n(z) \, .
\end{aligned}
\end{align}
The first two terms cancel each other. In the third term we use $\Box \, C(x-y) = - \d(x-y)$. It is then easy to see that
\begin{align}
\begin{aligned}
\pa_\m A_\m^{\prime \, a}(x) \big\vert_{\cO(g^3)} 
&=  \frac{g^3}{3} f^{abc} f^{bde} f^{dmn} \int \md y \ \md z  \ \bigg\{  \\
&\hspace{2cm}  + \pa_\r \left( A_\r^c(x)  C(x-y) \right)   A_\l^e(y) \pa_\s C(y-z) A_\l^m(z) A_\s^n(z) \\
&\hspace{2cm} - \pa_\m \left( A_\m^c(x) C(x-y) \right) A_\r^e(y) \pa_\l C(y-z) A_\r^m(z) A_\l^n(z) \bigg\}  \\
&=\; 0 \, .
\end{aligned}
\end{align}

\subsection{Free action}
By the first statement in the main theorem the transformed gauge field must satisfy
\begin{align}
\frac{1}{2} \int \md x \ A_\m^{\prime \, a}(x) \left(- \Box \, \d_{\m\n} + \pa_\m \pa_\n
\right) A_\n^{\prime \, a}(x) \,=\,  \frac{1}{4} \int \md x \ F_{\m \n}^a(x) F_{\m \n}^a(x)  
\,+ \, \cO(g^4) \, . 
\label{eq:FreeAction}
\end{align}
We stress that, unlike the matching of determinants,
the fulfillment of this condition will turn out {\em not to depend on the dimension 
$D$}. Because of the invariance of the Landau gauge function, we can ignore the 
second term on the l.h.s. and the corresponding term on the r.h.s. of this equation. The calculation, up to $\cO(g^2)$, is presented in detail in \cite{ANPP}. At third  order, \eqref{eq:FreeAction} 
has two contributions which must cancel each other:
\begin{align}
0 &\stackrel{!}{=}  \int \md x \ \bigg( A_\m^{\prime \, a}(x) \big\vert_{\cO(g^3)} \, \Box \,  A_\m^{\prime \, a}(x) \big\vert_{\cO(g^0)} +  A_\m^{\prime \, a}(x) \big\vert_{\cO(g^2)}  \, \Box  \, A_\m^{\prime \, a}(x) \big\vert_{\cO(g^1)}\bigg)  \, .
\end{align}
To check this we collect all the terms to obtain
\begin{align}
\begin{aligned}
&\int \md x \ \bigg( A_\m^{\prime \, a}(x) \big\vert_{\cO(g^3)} \, \Box \,  A_\m^{\prime \, a}(x) \big\vert_{\cO(g^0)} +  A_\m^{\prime \, a}(x) \big\vert_{\cO(g^2)}  \, \Box  \, A_\m^{\prime \, a}(x) \big\vert_{\cO(g^1)}\bigg)   \\
&\q= - \, \frac{g^3}{2} f^{a b c} f^{b d e} f^{c m n} \int \md x \ \md y \ \md z \ \md w \ \pa_\r C(x-y)  \\
&\hspace{3cm}  \times  \pa_\l C(y-z)  A_\l^d(z) A_\s^e(z) \pa_{[\r} C(y-w)  A_\m^m(w) A_{\s]}^n(w)  \,  \Box \, A_\m^a(x)  \\
&\qq + \, g^3 f^{a b c} f^{b d e} f^{d m n}  \int \md x \ \md y \ \md z \ \md w \ \pa_\r C(x-y) A_\l^c(y) \  \bigg\{  \\
&\hspace{2cm} + 2 \, \pa_{[\r} C(y-z)A_{\s]}^e(z)   \pa_{[\l} C(z-w)  A_\m^m(w) A_{\s]}^n(w) \,  \Box \, A_\m^a(x) \\
&\hspace{2cm}  - 2 \,  \pa_{[\l} C(y-z)A_{\s]}^e(z)   \pa_{[\r} C(z-w)  A_\m^m(w) A_{\s]}^n(w) \,  \Box \, A_\m^a(x) \\
&\hspace{2cm} - \pa_{\s} C(y-z)A_{\s}^e(z)   \pa_{[\r} C(z-w)  A_\m^m(w) A_{\l]}^n(w) \,  \Box \, A_\m^a(x)  \\
&\hspace{2cm}  - 2 \, \pa_{[\s} C(y-z)A_{\m]}^e(z)   \pa_{[\r} C(z-w)  A_\l^m(w) A_{\s]}^n(w) \,  \Box \, A_\m^a(x) \\
&\hspace{2cm}  +  \pa_{[\r} C(y-z) A_\m^e(z)  \pa_{ | \s |} C(z-w)  A_{\l ]}^m(w) A_\s^n(w) \,  \Box \, A_\m^a(x) \bigg\} \\
&\qq + \frac{g^3}{3} f^{abc} f^{bde} f^{dmn} \int \md x \  \md y \ \md z \ \md w \ \bigg\{ \\
&\hspace{2cm}
 +  2 \,  \pa_\r C(x-y) A_{[\r}^c(y) \pa_{\m]} C(y-z) A_\l^e(z) \pa_\s C(z-w) A_\l^m(w) A_\s^n(w) \,  \Box \,   A_\m^a(x)  \\
&\hspace{2cm}  - \pa_\m C(x-y) \, \pa_\r \left( A_\r^c(y)  C(y-z) \right) A_\l^e(z) \pa_\s C(z-w) A_\l^m(w) A_\s^n(w) \,  \Box \, A_\m^a(x) \bigg\} \\
&\qq - \, \frac{g^3}{3} f^{abc} f^{bde} f^{dmn} \int \md x \  \md y \ \md z    \\
&\hspace{3cm}  \times   A_\m^c(x) C(x-y) A_\r^e(y) \pa_\l C(y-z) A_\r^m(z) A_\l^n(z) \,  \Box \, A_\m^a(x) \\
&\qq  + \, \frac{3 g^3}{2}  f^{a b c} f^{b d e} \int \md x \  \md y \ \md z \  \md w \ \pa_\r C(x-y) A_\l^{c}(y) \pa_{[\m} C(y-z) A_\l^{d}(z) A_{\r ]}^{e}(z) \\
&\hspace{3cm}  \times \, \Box \, \left( f^{a m n}  \pa_\s C(x-w) A_\m^{m}(w) A_\s^{n}(w) \right) \, .
\end{aligned}
\end{align}
The general procedure to simplify this expression is rather straightforward. However, there are a few terms which require additional attention. The first step is to integrate each term by parts such that the Laplacian acts on the $C(x-y)$, which then simplifies to a $\d$-function and we obtain
\begin{align}
\begin{aligned}
&\int \md x \ \bigg( A_\m^{\prime \, a}(x) \big\vert_{\cO(g^3)} \, \Box \,  A_\m^{\prime \, a}(x) \big\vert_{\cO(g^0)} +  A_\m^{\prime \, a}(x) \big\vert_{\cO(g^2)}  \, \Box  \, A_\m^{\prime \, a}(x) \big\vert_{\cO(g^1)}\bigg)   \\
&\q= - \frac{g^3}{2} f^{a b c} f^{b d e} f^{c m n} \int \md x \  \md z \ \md w \ \pa_\r A_\m^a(x)   \\
&\hspace{3cm} \times  \pa_\l C(x-z)  A_\l^d(z) A_\s^e(z) \pa_{[\r} C(x-w)  A_\m^m(w) A_{\s]}^n(w)   \\
&\qq +  g^3 f^{a b c} f^{b d e} f^{d m n}  \int \md x \ \md z \ \md w \ \pa_\r A_\m^a(x)  A_\l^c(x) \  \bigg\{  \\
&\hspace{2cm} + 2 \, \pa_{[\r} C(x-z)A_{\s]}^e(z)   \pa_{[\l} C(z-w)  A_\m^m(w) A_{\s]}^n(w)  \\
&\hspace{2cm}  - 2 \,  \pa_{[\l} C(x-z)A_{\s]}^e(z)   \pa_{[\r} C(z-w)  A_\m^m(w) A_{\s]}^n(w)  \\
&\hspace{2cm}- \pa_{\s} C(x-z)A_{\s}^e(z)   \pa_{[\r} C(z-w)  A_\m^m(w) A_{\l]}^n(w)  \\
&\hspace{2cm}  - 2 \, \pa_{[\s} C(x-z)A_{\m]}^e(z)   \pa_{[\r} C(z-w)  A_\l^m(w) A_{\s]}^n(w) \\
&\hspace{2cm}  + \pa_{[\r} C(x-z) A_\m^e(z)  \pa_{ | \s |} C(z-w)  A_{\l ]}^m(w) A_\s^n(w)  \bigg\} \\
&\qq + \frac{g^3}{3} f^{abc} f^{bde} f^{dmn} \int \md x \ \md z \ \md w \ \bigg\{ \\
&\hspace{2cm} +  2 \,  \pa_\r A_\m^a(x)  A_{[\r}^c(x) \pa_{\m]} C(x-z) A_\l^e(z) \pa_\s C(z-w) A_\l^m(w) A_\s^n(w)  \\
&\hspace{2cm} 
  - \pa_\m A_\m^a(x)  \, \pa_\r \left( A_\r^c(x)  C(x-z) \right) A_\l^e(z) \pa_\s C(z-w) A_\l^m(w) A_\s^n(w)  \bigg\} \\
&\qq - \frac{g^3}{3} f^{abc} f^{bde} f^{dmn} \int \md x \  \md y \ \md z    \\
&\hspace{3cm}   \times   \Box \, A_\m^a(x)   A_\m^c(x) C(x-y) A_\r^e(y) \pa_\l C(y-z) A_\r^m(z) A_\l^n(z) \\
&\qq  + \frac{3 g^3}{2}  f^{a b c} f^{b d e} f^{a m n}  \int \md x \ \md z \  \md w \\
&\hspace{3cm}  \times   A_\l^{c}(x) \pa_{[\m} C(x-z) A_\l^{d}(z) A_{\r ]}^{e}(z)  \pa_\r \pa_\s C(x-w) A_\m^{m}(w) A_\s^{n}(w)  \, .
\end{aligned}
\end{align}
We notice that we can replace any $\pa_\r A_\m^a(x)  A_\l^c(x)$ by $\frac{1}{2} \pa_\r \left( A_\m^a(x)  A_\l^c(x) \right)$ if the full expression is symmetric under 
simultaneous exchange 
$a \leftrightarrow c$ and $\m \leftrightarrow \l$. This allows us to integrate the respective terms by parts again. After the integration the index contractions must be spelled out and most terms cancel. Subsequently, we obtain
\begin{align}
\begin{aligned}
&\int \md x \ \bigg( A_\m^{\prime \, a}(x) \big\vert_{\cO(g^3)} \, \Box \,  A_\m^{\prime \, a}(x) \big\vert_{\cO(g^0)} +  A_\m^{\prime \, a}(x) \big\vert_{\cO(g^2)}  \, \Box  \, A_\m^{\prime \, a}(x) \big\vert_{\cO(g^1)}\bigg)  \\
&\q=   \frac{g^3}{2} f^{a b c} f^{b d e} f^{d m n}  \int \md x  \ \md w \ 
 \pa_\r A_\m^a(x)  A_\l^c(x)  A_{\s}^e(x)   \pa_{[\l} C(x-w)  A_\m^m(w) A_{\s]}^n(w)  \\
&\qq +  \frac{g^3}{3} f^{abc} f^{bde} f^{dmn} \int \md x \ \md z \ \md w \ \bigg\{ \\
&\hspace{2cm} +  \pa_\r A_\m^a(x)  A_{\r}^c(x) \pa_{\m} C(x-z) A_\l^e(z) \pa_\s C(z-w) A_\l^m(w) A_\s^n(w)  \\
&\hspace{2cm} -  \pa_\r A_\m^a(x)  A_{\m }^c(x) \pa_{\r} C(x-z) A_\l^e(z) \pa_\s C(z-w) A_\l^m(w) A_\s^n(w)  \\
&\hspace{2cm}  - \pa_\m A_\m^a(x) \, \pa_\r \left( A_\r^c(x)  C(x-z) \right)A_\l^e(z) \pa_\s C(z-w) A_\l^m(w) A_\s^n(w) \bigg\} \\
&\qq -  \frac{g^3}{3} f^{abc} f^{bde} f^{dmn} \int \md x \  \md y \ \md z  \
  \Box \, A_\m^a(x)   A_\m^c(x) C(x-y) A_\r^e(y) \pa_\l C(y-z) A_\r^m(z) A_\l^n(z) \, .
\label{eq:FreeActionIntermediate}
\end{aligned}
\end{align}
The first term vanishes by the Jacobi identity, {\em i.e.}
\begin{align}
\begin{aligned}
&\frac{g^3}{2} f^{a b c} f^{b d e} f^{d m n}  \int \md x  \ \md w \  \pa_\r A_\m^a(x)  A_\l^c(x)  A_{\s}^e(x)   \pa_{[\l} C(x-w)  A_\m^m(w) A_{\s]}^n(w) \\
&\q = \, \frac{g^3}{6} \left( f^{a b c} f^{b d e} + f^{eba} f^{bdc} + f^{cbe} f^{bda} \right) f^{d m n}  \int \md x \ \md w \   \\
&\hspace{2cm} \times  \pa_\r A_\m^a(x)  A_\l^c(x)  A_{\s}^e(x)   \pa_{[\l} C(x-w)  A_\m^m(w) A_{\s]}^n(w) \\
&\q = \, 0 \, .
\end{aligned}
\end{align}
The second, third and fourth term in \eqref{eq:FreeActionIntermediate} can be integrated by parts and after removing the terms that are anti-symmetric under the exchange of two indices, we get
\begin{align}
\begin{aligned}
&\int \md x \ \bigg( A_\m^{\prime \, a}(x) \big\vert_{\cO(g^3)} \, \Box \,  A_\m^{\prime \, a}(x) \big\vert_{\cO(g^0)} +  A_\m^{\prime \, a}(x) \big\vert_{\cO(g^2)}  \, \Box  \, A_\m^{\prime \, a}(x) \big\vert_{\cO(g^1)}\bigg)   \\
&\q=  - \frac{g^3}{3} f^{abc} f^{bde} f^{dmn} \int \md x \ \md z \ \md w \ \bigg\{ \\
&\hspace{2cm} +  \pa_{\m} \pa_\r A_\m^a(x)  A_{\r}^c(x)  C(x-z) A_\l^e(z) \pa_\s C(z-w) A_\l^m(w) A_\s^n(w)  \\
&\hspace{2cm} -  \Box A_\m^a(x)  A_{\m }^c(x) C(x-z) A_\l^e(z) \pa_\s C(z-w) A_\l^m(w) A_\s^n(w)  \\
&\hspace{2cm}   - \pa_\m  \pa_\r A_\m^a(x)  A_\r^c(x)  C(x-z) A_\l^e(z) \pa_\s C(z-w) A_\l^m(w) A_\s^n(w)  \bigg\} \\
&\qq - \frac{g^3}{3} f^{abc} f^{bde} f^{dmn} \int \md x \  \md y \ \md z \,
       \Box \, A_\m^a(x)   A_\m^c(x) C(x-y) A_\r^e(y) \pa_\l C(y-z) A_\r^m(z) A_\l^n(z) \\
&\q= \; 0 \, .
\end{aligned}
\end{align}
Thus, the condition \eqref{eq:FreeAction} holds up to and including $\cO(g^3)$. It is worthwhile 
to point out here that the very existence of a non-local field transformation mapping one
local action to another local action is a remarkable fact in itself, independently of
supersymmetry (but in the absence of supersymmetry, locality would be spoilt
by the Jacobian).

%\newpage
\subsection{Jacobians, fermion and ghost determinants}
Finally, we need to perturbatively show that on the gauge surface the Jacobian determinant is equal to the product of the MSS and FP determinants. This is done order by order in $g$ by considering the logarithms of the determinants rather than the determinants themselves; since the
relevant checks up to $\cO(g^2)$ were already performed in \cite{Nic1,ANPP}, we
can here concentrate on the third order, {\em viz.}
\begin{align}
\log \det \left( \frac{\d A_\m^{\prime \, a}(x)}{\d A_\n^b(y)} \right) \bigg\vert_{\cO(g^3)} \,
\stackrel{!}{=} \; \log \Big(\D_{MSS}[A] \ \D_{FP}[A] \Big) \bigg\vert_{\cO(g^3)} \, .
\label{eq:JacobianFPMSSDet}
\end{align}
Of the three statements in subsection \ref{th:MainTheorem} this is the most complicated 
condition to verify. Moreover, it is the only condition that depends on the dimension of 
our field theory and will impose the constraint (\ref{rD})  on the latter.

\vskip 0.3cm
\noindent The ghost determinant is computed from the functional matrix
\begin{align}\label{bfX}
\mathbf{X}^{ab}(x,y;A) = g \, f^{abc} C(x-y) A_\m^c(y) \pa_\m^y\, ,
\end{align}
using the well-known equation
\begin{align}\label{Fdet}
\log \det \left(1 - \mathbf{X} \right) = \tr \, \log \left(1 - \mathbf{X} \right) \, .
\end{align}
Up to $\cO(g^3)$ this yields
\begin{align}
\begin{aligned}\label{bfX1}
 \log \det \left(1- \mathbf{X} \right)  &= \frac{1}{2} \, n \, g^2 \, \int \md x \ \md y \ \pa_\m C(x-y) A_\n^a(y) \pa_\n C(y-x) A_\m^a(x) \\
&\q + \frac{1}{3} \, g^3 \,  f^{a d m} f^{b e m} f^{c d e} \,  \int \md x \ \md y \ \md z \\
&\qq \times \pa_\m C(x-y) A_\n^b(y) \pa_\n C(y-z) A_\r^c(z)  \pa_\r C(z-x) A_\m^a(x) \\
&\q + \cO(g^4) \, .
\end{aligned}
\end{align}
Observe that in (\ref{bfX}) there are no terms containing 
$\pa^\mu A_\mu^c$: these do not contribute, as one may directly verify by integrating 
by parts and reading (\ref{bfX1}) `backwards'. Therefore the gauge condition (\ref{Landau})
is not required for the evaluation of the ghost determinant.

\vskip 0.3cm
\noindent The relevant kernel for the MSS determinant is
\begin{align}
\mathbf{Y}_{\a \b}^{ab}(x,y;A) =  g \, f^{abc} \pa_\r C(x-y) \left( \c^\r \c^\l \right)_{\a \b} A_\l^c(y) \, .
\end{align}
For the correct counting of physical fermionic degrees of 
freedom we must include an extra factor of $\frac12$ in the expansion (\ref{Fdet}) 
(technically due to the Majorana or Weyl condition when performing the 
Berezin integration) and get
\begin{align}
\begin{aligned}
\frac{1}{2} \, \log \det \left(1- \mathbf{Y} \right) 
&= \frac{1}{4} \, n \, g^2 \, \tr (\c^\r \c^\l \c^\s \c^\a) \, \int \md x \ \md y \ \pa_\r C(x-y) A_\l^a(y) \pa_\s C(y-x) A_\a^a(x) \\
& \q + \frac{1}{6} \, g^3 \,  f^{a d m} f^{b e m} f^{c d e} \,   \tr (\c^\r \c^\l \c^\s \c^\a \c^\b \c^\tau) \int \md x \ \md y \  \md z \  \\
&\qq \times \pa_\r C(x-y) A_\l^b(y) \pa_\s C(y-z) A_\a^c(z) \pa_\b C(z-x) A_\tau^a(x) \\
&\q + \cO(g^4) \, .
\end{aligned}
\label{eq:MSSDet}
\end{align}
For both determinants there is no contribution at $\cO(g)$ and we have simplified the results at $\cO(g^2)$ by using (\ref{ff}). Taking the trace in \eqref{eq:MSSDet} and multiplying the two determinants subsequently yields the right hand side of \eqref{eq:JacobianFPMSSDet}
\begin{align}
\begin{aligned}
\log &\left(\D_{MSS}[A] \ \D_{FP}[A] \right) \big\vert_{\cO(g^3)}  = f^{a d m} f^{b e m} f^{c d e}  \ \int \md x \ \md y \ \md z \ \bigg\{ \\
&\hspace{2cm} \color{red} -  r \,     \pa_\m C(x-y)   A_\m^b(y)  \pa_\r C(y-z) A_\l^c(z)   \pa_\r C(z-x)  A_\l^a(x)  \\ 
&\hspace{2cm} \color{green!20!blue} + \frac{r+1}{3} \,   \pa_\m C(x-y) A_\r^b(y) \pa_\r C(y-z) A_\l^c(z)  \pa_\l C(z-x) A_\m^a(x) \\ 
&\hspace{2cm}  \color{orange!80!red}+ \frac{r}{2} \,   \pa_\m C(x-y)  A_\r^b(y) \pa_\r C(y-z)  A_\m^c(z)  \pa_\l C(z-x) A_\l^a(x) \\ 
&\hspace{2cm} \color{green!50!black} - \frac{r}{6} \,  \pa_\m C(x-y) A_\r^b(y) \pa_\l C(y-z) A_\m^c(z)   \pa_\r C(z-x) A_\l^a(x) \\ 
&\hspace{2cm} \color{violet} + \frac{r}{2}  \,   \pa_\m C(x-y) A_\r^b(y) \pa_\l C(y-z)  A_\m^c(z)  \pa_\l C(z-x)  A_\r^a(x)  \color{black} \bigg\} \, .
\end{aligned}
\end{align}
We thus end up with a total of five independent structures; we use color coding to help us identify the corresponding terms in the Jacobian determinant. \\
At $\cO(g^3)$ the logarithm of the Jacobian determinant schematically consists of three terms  
\begin{align}
\begin{aligned}
\log \det \left( \frac{\d A_\mu^{\prime \, a}(x)}{\d A_\n^b(y)} \right) \bigg\vert_{\cO(g^3)} 
&=\; \tr \left[ \frac{\d A^\prime}{\d A} \bigg\vert_{\cO(g^3)} \right]  \,- \,
\left( 2 \cdot \frac{1}{2} \right) \tr \left[ \frac{\d A^\prime}{\d A}  \bigg\vert_{\cO(g^2)} 
\frac{\d A^\prime}{\d A} \bigg\vert_{\cO(g^1)} \right] \\[1mm]
&\hspace{2cm}  +  \, \frac{1}{3} \, \tr \left[ \frac{\d A^{\prime}}{\d A} \bigg\vert_{\cO(g^1)} 
\frac{\d A^{\prime}}{\d A} \bigg\vert_{\cO(g^1)} 
\frac{\d A^{\prime}}{\d A} \bigg\vert_{\cO(g^1)} \right] \, .
\end{aligned}
\end{align}
and the final trace is done by setting $\m=\n, a=b, x=y$ and integrating over $x$. The computation is straightforward and we find
\begin{align}
\begin{aligned}
&\frac{1}{3} \, \tr \left[ \frac{\d A_\m^{\prime \, a}(x)}{\d A_\n^b(y)} \bigg\vert_{\cO(g^1)} \frac{\d A_\m^{\prime \, a}(x)}{\d A_\n^b(y)} \bigg\vert_{\cO(g^1)} \frac{\d A_\m^{\prime \, a}(x)}{\d A_\n^b(y)} \bigg\vert_{\cO(g^1)} \right] \\
&\q =f^{a d m} f^{b e m} f^{c d e}  \ \int \md x \ \md y \ \md z \ \bigg\{ \\
&\hspace{2cm} \color{green!20!blue} - \frac{3-D}{3} \,   \pa_\m C(x-y) A_\r^b(y) \pa_\r C(y-z) A_\l^c(z)  \pa_\l C(z-x) A_\m^a(x) \\
&\hspace{2cm}\color{orange!80!red} +  \pa_\m C(x-y)  A_\r^b(y) \pa_\r C(y-z)  A_\m^c(z)  \pa_\l C(z-x) A_\l^a(x)  \\ 
&\hspace{2cm} \color{green!50!black} - \frac{1}{3} \,   \pa_\m C(x-y) A_\r^b(y) \pa_\l C(y-z) A_\m^c(z)   \pa_\r C(z-x) A_\l^a(x) \color{black}  \bigg\} \, .
\end{aligned}
\end{align}
The second term gives
\begin{align}
\begin{aligned}
-& \left( 2 \cdot \frac{1}{2} \right) \tr \left[ \frac{\d A_\m^{\prime \, a}(x)}{\d A_\n^b(y)} \bigg\vert_{\cO(g^2)} \frac{\d A_\m^{\prime \, a}(x)}{\d A_\n^b(y)} \bigg\vert_{\cO(g^1)} \right] \\
&\q = f^{a d m} f^{b e m} f^{c d e}  \ \int \md x \ \md y \ \md z \ \bigg\{ \\
&\hspace{2cm}  \color{red} + \frac{1-D}{2}   \,  \pa_\m C(x-y)   A_\m^b(y)  \pa_\r C(y-z) A_\l^c(z)   \pa_\r C(z-x)  A_\l^a(x)  \\ 
&\hspace{2cm}  \color{green!20!blue} + \frac{1}{2}  \,   \pa_\m C(x-y) A_\r^b(y) \pa_\r C(y-z) A_\l^c(z)  \pa_\l C(z-x) A_\m^a(x) \\ 
&\hspace{2cm} \color{orange!80!red} - \frac{3-D}{2} \,  \pa_\m C(x-y)  A_\r^b(y) \pa_\r C(y-z)  A_\m^c(z)  \pa_\l C(z-x) A_\l^a(x)\\ 
&\hspace{2cm} \color{violet} + \frac{1}{2} \,  \pa_\m C(x-y) A_\r^b(y) \pa_\l C(y-z)  A_\m^c(z)  \pa_\l C(z-x)  A_\r^a(x)    \color{black} \bigg\} \, .
\end{aligned}
\end{align}
Finally, the first term gives
\begin{align}
\begin{aligned}
&\tr \left[ \frac{\d A_\m^{\prime \, a}(x)}{\d A_\n^b(y)} \bigg\vert_{\cO(g^3)} \right]  \\
&\q =f^{a d m} f^{b e m} f^{c d e}  \ \int \md x \ \md y \ \md z \ \bigg\{ \\
&\hspace{2cm}\color{red} + \frac{7-3D}{2}   \,  \pa_\m C(x-y)   A_\m^b(y)  \pa_\r C(y-z) A_\l^c(z)   \pa_\r C(z-x)  A_\l^a(x)  \\ 
&\hspace{2cm}\color{green!20!blue} - \frac{3-2D}{6} \,   \pa_\m C(x-y) A_\r^b(y) \pa_\r C(y-z) A_\l^c(z)  \pa_\l C(z-x) A_\m^a(x) \\ 
&\hspace{2cm} \color{orange!80!red} - \frac{3-D}{2}    \,  \pa_\m C(x-y)  A_\r^b(y) \pa_\r C(y-z)  A_\m^c(z)  \pa_\l C(z-x) A_\l^a(x)\\  
&\hspace{2cm}  \color{green!50!black} + \frac{3-D}{3}  \,  \pa_\m C(x-y) A_\r^b(y) \pa_\l C(y-z) A_\m^c(z)   \pa_\r C(z-x) A_\l^a(x) \\ 
&\hspace{2cm} \color{violet}-  \frac{5-2D}{2}  \,  \pa_\m C(x-y) A_\r^b(y) \pa_\l C(y-z)  A_\m^c(z)  \pa_\l C(z-x)  A_\r^a(x)   \color{black} \bigg\} \\ 
&\qq - \frac{2}{3} \,  f^{aem} f^{bde} f^{cdm}  \ \int \md x \ \md y \  A_\m^b(x) A_\r^c(x) C(x-y) \pa_\r C(x-y) A_\m^a(y) \\
&\qq + \frac{1}{3} \, f^{adm} f^{bce} f^{dem} \int  \md x \ \md y \ \md z  \  A_\m^a(x) \left(  \pa_\r C(x-y) \right)^2  \pa_\l C(y-z) A_\l^b(z) A_\m^c(z)  \\
&\q  \q- \frac{1}{3} \,  f^{adm} f^{bce} f^{dem} \int  \md x \ \md y \ C(0) A_\m^a(x)   \pa_\r C(x-y) A_\r^b(y) A_\m^c(y) \, .
\end{aligned}
\label{eq:JacobianFirstTerm}
\end{align}
There are two special features about this part of the Jacobian determinant. First, we have to use 
the gauge condition $G^a(A) \equiv \partial^\mu A_\mu^a = 0$ to eliminate two terms. Secondly, 
we find terms which do not match any of the five structures from the fermion and ghost determinants and hence must cancel among themselves. 
However, before addressing those terms, let us first analyze the color coded terms. Imposing the equality \eqref{eq:JacobianFPMSSDet} yields the following conditions
\begin{align}
\begin{aligned}
\color{red} -r &\color{red} \;\;=\;  \frac{1-D}{2} + \frac{7-3D}{2} \;=\; 4 - 2D \\
\color{green!20!blue} \frac{r+1}{3} &\color{green!20!blue}
\;\;=\; - \frac{3-D}{3} + \frac{1}{2} - \frac{3-2D}{6} \;=\; \frac{2D-3}{3} \\
\color{orange!80!red} \frac{r}{2} & \color{orange!80!red}
\;\;=\; 1 - \frac{3-D}{2} - \frac{3-D}{2} \;=\; D-2 \\
\color{green!50!black} - \frac{r}{6} & \color{green!50!black} 
\;\;=\; - \frac{1}{3} + \frac{3-D}{3} \;=\; \frac{2-D}{3} \\
\color{violet} \frac{r}{2} & \color{violet} 
\;\;=\;  \frac{1}{2} - \frac{5-2D}{2} \;=\; D -2 \ .
\end{aligned}
\end{align}
Happily, all five equations are satisfied with $r = 2(D-2)$, so we recover the result (\ref{rD})
\begin{align}
D = 3, 4, 6, 10 \qq \Longleftrightarrow \qq r = 2, 4, 8, 16 \,  ,
\end{align}
thus extending the result of \cite{ANPP} to cubic order. It remains to be shown that the remaining (black) 
terms from \eqref{eq:JacobianFirstTerm} vanish. Using the Jacobi identity in the first term 
and $f^{abc} f^{abd} = n \,  \d^{cd}$ in the latter two yields
\begin{align}
\begin{aligned}
& - \frac{n}{3}f^{abc}   \int \md x \ \md y \  A_\m^b(x) A_\r^c(x) C(x-y) \pa_\r C(x-y) A_\m^a(y) \\
&+ \frac{n}{3}f^{abc} \int  \md x \ \md y \ \md z  \  A_\m^a(x) \left(  \pa_\r C(x-y) \right)^2  \pa_\l C(y-z) A_\l^b(z) A_\m^c(z)  \\
&- \frac{n}{3}f^{abc}  \int  \md x \ \md y \ C(0) A_\m^a(x)   \pa_\r C(x-y) A_\r^b(y) A_\m^c(y) \, .
\end{aligned}
\end{align} 
The second term is rewritten using the identity
\begin{align}
\Box \left( C^2(x-y) \right) \,=\, 
     - 2 \, C(0)\d(x-y) \,+\, 2 \,  \pa_\r C(x-y) \pa_\r C(x-y)  \, , 
\end{align}
with a formally divergent piece $C(0)$ which can be appropriately regulated. This simplifies the expression above to
\begin{align}
\begin{aligned}
&\q -\frac{n}{3} \, f^{abc}  \int \md x \ \md y \    A_\m^b(x) A_\r^c(x) C(x-y) \pa_\r C(x-y) A_\m^a(y)  \\
&\q + \frac{n}{3} \,  f^{abc}  \int  \md x \ \md y \ \md z  \    A_\m^a(x)   C(0) \d(x-y)  \pa_\r C(y-z) A_\r^b(z) A_\m^c(z)  \\
&\q + \frac{n}{6} \,   f^{abc}  \int  \md x \ \md y \ \md z  \    A_\m^a(x) \, \Box \, \left( C^2(x-y)  \right)   \pa_\r C(y-z) A_\r^b(z) A_\m^c(z)  \\
&\q-\frac{n}{3} \, f^{abc}  \int  \md x \ \md y \    C(0) A_\m^a(x) \pa_\r C(x-y)   A_\r^b(y) A_\m^c(y)\, .
\end{aligned}
\end{align} 
Subsequent integration by parts produces
\begin{align}
\begin{aligned}
&\q  \frac{n}{3} \, f^{abc}  \int \md x \ \md y \  A_\m^b(x) A_\r^c(x) C(x-y) \pa_\r C(x-y)  A_\m^a(y)   \\
&\q +  \frac{n}{3} \,  f^{abc}  \int  \md x \ \md y \  C(0)  A_\m^a(x) \pa_\r C(x-y)   A_\r^b(y) A_\m^c(y)  \\
&\q -  \frac{n}{6}  \, f^{abc}  \int  \md x \ \md y \ \md z  \    A_\m^a(x)   \pa_\r   \left( C^2(x-y)  \right) \, \Box\,  C(y-z) A_\r^b(z) A_\m^c(z)  \\
&\q  - \frac{n}{3} \, f^{abc}  \int  \md x \ \md y \    C(0)  A_\m^a(x)  \pa_\r C(x-y) A_\r^b(y) A_\m^c(y) \\
&=  \frac{n}{3} \, f^{abc}  \ \int \md x \ \md y \   A_\m^b(x) A_\r^c(x) C(x-y) \pa_\r C(x-y)  A_\m^a(y)  \\
&\q +  \frac{n}{6} \,   f^{abc}  \int  \md x \ \md y   \    A_\m^a(x)   \pa_\r   \left( C^2(x-y)  \right) A_\r^b(y) A_\m^c(y)  \\
&= 0 \, .
\end{aligned}
\end{align} 
Thus, \eqref{eq:JacobianFPMSSDet} is satisfied. Let us note that, unlike for the
$\cO(g^2)$ computation, we had to make use of the Landau gauge condition (\ref{Landau})
to achieve this equality. This feature, which arises only from $\cO(g^3)$ onwards, 
is entirely due to the $g$-dependence of the ghost propagator in (\ref{eq:DelGhostProp}).

\newpage
\section{Beyond the third order: graphical representation}

It is not difficult to present the full perturbative expansion of the $\mathcal{R}$ operator. 
To streamline the notation, we suppress the color indices and position variables and integrations,
but understand all objects to be multiplied as color matrices or vectors and 
by convoluting integration kernels with $A$ insertions. Concretely, abbreviating
\begin{align}
\begin{aligned}
&(A_\m)^{ab}(x) = f^{amb}  A^m_\m(x) \qq\Rightarrow\qq
\slashed{A} = \c^\m  A_\m\, ,\q
\pa{\cdot}A = \pa^\m A_\m\, , \\
&(\slashed{A}^2)^a(x) = f^{abc} \c^{\r\l} A^b_\r(x) A^c_\l(x)
\end{aligned}
\end{align}
and expanding
\begin{align}
\begin{aligned}
S &\ =\ S_0 - g\,S_0 \slashed{A}\,S \ =\ S_0\ \sum_{\ell=0}^\infty \bigl(-g\,\slashed{A}\,S_0\bigr)^\ell\, ,\\
G &\ =\ 
G_0 -g\,G_0\,\pa{\cdot}A\,G \ =\ G_0\ \sum_{k=0}^\infty \bigl(-g\,\pa{\cdot}A\,G_0\bigr)^k\, ,
\end{aligned}
\label{eq:fullprops}
\end{align}
where derivatives act on everything on their right, we may write
\begin{align}
\begin{aligned}
\mathcal{R} &\ =\  \overleftarrow{\frac{\md}{\md g}} 
- \frac{1}{2r} \overleftarrow{\frac{\d}{\d A_\m}} \Pi_\m^{\ \s}
\Bigl\{ \d_{\s\n} - g A_\s G_0 \ \sum_{k=0}^\infty \bigl(-g\,\pa{\cdot}A\,G_0\bigr)^k \pa_\n \Bigr\} \times
\tr \Bigl\{ \c^\n S_0 \ \sum_{\ell=0}^\infty \bigl(-g \slashed{A}\,S_0\bigr)^\ell \slashed{A}^2\Bigr\} \\[4pt]
&\ =\ \overleftarrow{\frac{\md}{\md g}} 
- \frac{1}{2r} \overleftarrow{\frac{\d}{\d A_\m}} \Pi_{\m\n} \tr \biggl\{ 
\Bigl[ \c^\n S_0 - g A^\n G_0\ \sum_{k=0}^\infty \bigl(-g\,\pa{\cdot}A\,G_0\bigr)^k \Bigr]
\sum_{\ell=0}^\infty \bigl(-g \slashed{A}\,S_0\bigr)^\ell \slashed{A}^2\biggr\} \\[4pt]
&\ =\ \overleftarrow{\frac{\md}{\md g}} 
- \frac{1}{2r} \overleftarrow{\frac{\d}{\d A_\m}}  \tr \biggl\{ \Bigl[
\c_{\m\n}\pa^\n\,+\, 2g\,\delta_{\m\n}^{\a\b}\pa_\a G_0\,\pa^\n\! A_\b 
\sum_{k=0}^\infty \bigl(-g\,G_0 \pa{\cdot}A\bigr)^k \Bigr] G_0
\sum_{\ell=0}^\infty \bigl(-g \slashed{A}\,S_0\bigr)^\ell \slashed{A}^2\biggr\}
\, ,
\end{aligned}
\end{align}
where we now let the variation with respect to $g$ and $A$ act {\em to the left\/} 
in order to conform with the implicit color and position ordering in these expressions. 
In the last line, we have expanded the abelian transversal projector~$\Pi$ and simplified
\begin{align}
\begin{aligned}
\Pi_{\m\n} \c^\n S_0 \ &=\ \c_\m S_0 - \pa_\m G_0\,\pa_\n\c^\n S_0 
\ =\ \c_\m\slashed{\pa}G_0 - \pa_\m G_0 \ =\ (\c_\m\c_\n-\delta_{\m\n})\pa^\n G_0
\ =\ \c_{\m\n}\,\pa^\n G_0\, ,\\[4pt]
\Pi_{\m\n} A^\n G_0 &= A_\m G_0 - \pa_\m G_0\,\pa_\n A^\n G_0 
= (\Box G_0  A_\m - \pa_\m \pa^\n G_0 A_\n) G_0
\ =\ -2\delta_{\m\n}^{\a\b} \pa_\a G_0\,\pa^\n\!A_\b G_0\, .
\end{aligned}
\end{align}

\vskip 0.3cm
\noindent 
The individual terms of the expansion in powers of~$g$ allow for a simple representation 
in terms of Feynman diagrams, built from free fermion and ghost propagators dressed with photon insertions.
They have the graphical structure of linear trees mostly starting with a modified double ghost line
$2\delta_{\m\n}^{\a\b} \pa_\a G_0\,\pa^\n\!A_\b G_0$ 
and ending in a ``double photon emission"~$\sfrac12\slashed{A}^2$. 
In between, one finds $k$ further ghost lines ($k=0,1,2,\ldots$) 
followed by $\ell$ fermion lines ($\ell=0,1,2,\ldots$), 
separated by the appropriate photon emission insertions. 
Since the initial $A$-variation reduces the power of~$A$ by one, 
such a term contributes to $\cO(g^{k+\ell+1}A^{k+\ell+2})$. 
In addition, there are the ghost-free linear trees, which are rooted in a modified fermion line
$\c_{\m\n}\,\pa^\n G_0$, comprise $\ell$ additional fermion lines and also terminate in~$\sfrac12\slashed{A}^2$. 
They contribute to $\cO(g^{\ell}A^{\ell+1})$. 
For a given order $\cO(g^n A^{n+1})$ then, one encounters $n{+}1$ linear trees of length $n{+}1$:
a single ghost-free one of structure $S_0^{n+1}$
and $n$ mixed ones of structure $G_0^2S_0^{n-1},\,G_0^3S_0^{n-2},\,\ldots,\,G_0^nS_0,\,G_0^{n+1}$.
The last of these always vanishes due to $\tr\slashed{A}^2=0$.
The number of gamma matrices in the traces are $2n{+}4,\,2n,\,2n{-}2,\,2n{-}4,\,\ldots,\,4,\,2,$ respectively.
Below we illustrate the graphical expansion of $\mathcal{R}$ up to $\cO(g^4)$:

\setlength{\unitlength}{1cm}
\thicklines
\hspace{5mm}
\begin{picture}(15.5,12)
%\put(-1,0){\line(0,1){12}}
%\put(-1,0){\line(1,0){16.5}}
%\put(-1,12){\line(1,0){16.5}}
%\put(15.5,0){\line(0,1){12}}
%
\put(-1,10.91){$\mathcal{R}\q=\q\overleftarrow{\frac{\md}{\md g}} 
\q+\q R_1 \q+\q g\,R_2 \q+\q g^2 R_3 \q+\q g^3 R_4 \q+\q \ldots$}
\put(-0.2,8.91){$=\q\overleftarrow{\frac{\md}{\md g}} \q + \q$}
\put(3.5,9){\vector(-1,0){1.2}}
\put(3.54,9.03){\oval(0.1,0.1)[tl]}     % startNE
\put(3.59,9.15){\oval(0.1,0.1)[br]}    % +0.05 +0.12
\put(3.69,9.18){\oval(0.1,0.1)[tl]}      % +0.1 +0.03
\put(3.74,9.3){\oval(0.1,0.1)[br]}      % +0.05 +0.12
\put(3.54,9.0){\oval(0.1,0.1)[bl]}      % startSE
\put(3.59,8.88){\oval(0.1,0.1)[tr]}     % +0.05 -0.12
\put(3.69,8.85){\oval(0.1,0.1)[bl]}     % +0.1 -0.03
\put(3.74,8.73){\oval(0.1,0.1)[tr]}     % +0.05 -0.12
\put(0.5,6.91){$+\q g$}
\put(3.5,7){\vector(-1,0){1.7}}
\put(2.7,6.99){\oval(0.1,0.1)[bl]}       % startS
\put(2.74,6.87){\oval(0.1,0.1)[tr]}     % +0.04 -0.12
\put(2.74,6.84){\oval(0.1,0.1)[br]}    % 0 -0.03
\put(2.7,6.72){\oval(0.1,0.1)[tl]}        % -0.04 -0.12
\put(2.7,6.69){\oval(0.1,0.1)[bl]}       % 0 -0.03
\put(2.74,6.57){\oval(0.1,0.1)[tr]}     % +0.04 -0.12
\put(2.74,6.54){\oval(0.1,0.1)[br]}    % 0 -0.03
\put(2.7,6.42){\oval(0.1,0.1)[tl]}       % -0.04 -0.12
\put(3.54,7.03){\oval(0.1,0.1)[tl]}     % startNE
\put(3.59,7.15){\oval(0.1,0.1)[br]}    % +0.05 +0.12
\put(3.69,7.18){\oval(0.1,0.1)[tl]}      % +0.1 +0.03
\put(3.74,7.3){\oval(0.1,0.1)[br]}      % +0.05 +0.12
\put(3.54,7.0){\oval(0.1,0.1)[bl]}      % startSE
\put(3.59,6.88){\oval(0.1,0.1)[tr]}     % +0.05 -0.12
\put(3.69,6.85){\oval(0.1,0.1)[bl]}     % +0.1 -0.03
\put(3.74,6.73){\oval(0.1,0.1)[tr]}     % +0.05 -0.12
\put(0.5,4.91){$+\q g^2$}
\put(4.3,5){\vector(-1,0){2.5}}
\put(2.7,4.99){\oval(0.1,0.1)[bl]}      % startS
\put(2.74,4.87){\oval(0.1,0.1)[tr]}     % +0.04 -0.12
\put(2.74,4.84){\oval(0.1,0.1)[br]}    % 0 -0.03
\put(2.7,4.72){\oval(0.1,0.1)[tl]}        % -0.04 -0.12
\put(2.7,4.69){\oval(0.1,0.1)[bl]}       % 0 -0.03
\put(2.74,4.57){\oval(0.1,0.1)[tr]}     % +0.04 -0.12
\put(2.74,4.54){\oval(0.1,0.1)[br]}    % 0 -0.03
\put(2.7,4.42){\oval(0.1,0.1)[tl]}       % -0.04 -0.12
\put(3.5,4.99){\oval(0.1,0.1)[bl]}      % startS
\put(3.54,4.87){\oval(0.1,0.1)[tr]}     % +0.04 -0.12
\put(3.54,4.84){\oval(0.1,0.1)[br]}    % 0 -0.03
\put(3.5,4.72){\oval(0.1,0.1)[tl]}        % -0.04 -0.12
\put(3.5,4.69){\oval(0.1,0.1)[bl]}       % 0 -0.03
\put(3.54,4.57){\oval(0.1,0.1)[tr]}     % +0.04 -0.12
\put(3.54,4.54){\oval(0.1,0.1)[br]}    % 0 -0.03
\put(3.5,4.42){\oval(0.1,0.1)[tl]}       % -0.04 -0.12
\put(4.34,5.03){\oval(0.1,0.1)[tl]}     % startNE
\put(4.39,5.15){\oval(0.1,0.1)[br]}    % +0.05 +0.12
\put(4.49,5.18){\oval(0.1,0.1)[tl]}      % +0.1 +0.03
\put(4.54,5.3){\oval(0.1,0.1)[br]}      % +0.05 +0.12
\put(4.34,5.0){\oval(0.1,0.1)[bl]}      % startSE
\put(4.39,4.88){\oval(0.1,0.1)[tr]}     % +0.05 -0.12
\put(4.49,4.85){\oval(0.1,0.1)[bl]}     % +0.1 -0.03
\put(4.54,4.73){\oval(0.1,0.1)[tr]}     % +0.05 -0.12
\put(5.1,4.91){$+\q g^2$}
\put(6.5,5){\vector(-1,0){0.1}}
\multiput(6.5,5)(0.2,0){8}{\line(1,0){0.1}}
\put(8.1,5){\line(1,0){0.8}}
\put(7.4,4.99){\oval(0.1,0.1)[bl]}      % startS
\put(7.44,4.87){\oval(0.1,0.1)[tr]}     % +0.04 -0.12
\put(7.44,4.84){\oval(0.1,0.1)[br]}    % 0 -0.03
\put(7.4,4.72){\oval(0.1,0.1)[tl]}        % -0.04 -0.12
\put(7.4,4.69){\oval(0.1,0.1)[bl]}       % 0 -0.03
\put(7.44,4.57){\oval(0.1,0.1)[tr]}     % +0.04 -0.12
\put(7.44,4.54){\oval(0.1,0.1)[br]}    % 0 -0.03
\put(7.4,4.42){\oval(0.1,0.1)[tl]}       % -0.04 -0.12
\put(8.2,4.99){\oval(0.1,0.1)[bl]}      % startS
\put(8.24,4.87){\oval(0.1,0.1)[tr]}     % +0.04 -0.12
\put(8.24,4.84){\oval(0.1,0.1)[br]}    % 0 -0.03
\put(8.2,4.72){\oval(0.1,0.1)[tl]}        % -0.04 -0.12
\put(8.2,4.69){\oval(0.1,0.1)[bl]}       % 0 -0.03
\put(8.24,4.57){\oval(0.1,0.1)[tr]}     % +0.04 -0.12
\put(8.24,4.54){\oval(0.1,0.1)[br]}    % 0 -0.03
\put(8.2,4.42){\oval(0.1,0.1)[tl]}       % -0.04 -0.12
\put(8.94,5.03){\oval(0.1,0.1)[tl]}     % startNE
\put(8.99,5.15){\oval(0.1,0.1)[br]}    % +0.05 +0.12
\put(9.09,5.18){\oval(0.1,0.1)[tl]}      % +0.1 +0.03
\put(9.14,5.3){\oval(0.1,0.1)[br]}      % +0.05 +0.12
\put(8.94,5.0){\oval(0.1,0.1)[bl]}      % startSE
\put(8.99,4.88){\oval(0.1,0.1)[tr]}     % +0.05 -0.12
\put(9.09,4.85){\oval(0.1,0.1)[bl]}     % +0.1 -0.03
\put(9.14,4.73){\oval(0.1,0.1)[tr]}     % +0.05 -0.12
\put(0.5,2.91){$+\q g^3$}
\put(4.7,3){\vector(-1,0){2.9}}
\put(2.6,2.99){\oval(0.1,0.1)[bl]}      % startS
\put(2.64,2.87){\oval(0.1,0.1)[tr]}     % +0.04 -0.12
\put(2.64,2.84){\oval(0.1,0.1)[br]}    % 0 -0.03
\put(2.6,2.72){\oval(0.1,0.1)[tl]}        % -0.04 -0.12
\put(2.6,2.69){\oval(0.1,0.1)[bl]}       % 0 -0.03
\put(2.64,2.57){\oval(0.1,0.1)[tr]}     % +0.04 -0.12
\put(2.64,2.54){\oval(0.1,0.1)[br]}    % 0 -0.03
\put(2.6,2.42){\oval(0.1,0.1)[tl]}       % -0.04 -0.12
\put(3.3,2.99){\oval(0.1,0.1)[bl]}      % startS
\put(3.34,2.87){\oval(0.1,0.1)[tr]}     % +0.04 -0.12
\put(3.34,2.84){\oval(0.1,0.1)[br]}    % 0 -0.03
\put(3.3,2.72){\oval(0.1,0.1)[tl]}        % -0.04 -0.12
\put(3.3,2.69){\oval(0.1,0.1)[bl]}       % 0 -0.03
\put(3.34,2.57){\oval(0.1,0.1)[tr]}     % +0.04 -0.12
\put(3.34,2.54){\oval(0.1,0.1)[br]}    % 0 -0.03
\put(3.3,2.42){\oval(0.1,0.1)[tl]}       % -0.04 -0.12
\put(4,2.99){\oval(0.1,0.1)[bl]}      % startS
\put(4.04,2.87){\oval(0.1,0.1)[tr]}     % +0.04 -0.12
\put(4.04,2.84){\oval(0.1,0.1)[br]}    % 0 -0.03
\put(4,2.72){\oval(0.1,0.1)[tl]}        % -0.04 -0.12
\put(4,2.69){\oval(0.1,0.1)[bl]}       % 0 -0.03
\put(4.04,2.57){\oval(0.1,0.1)[tr]}     % +0.04 -0.12
\put(4.04,2.54){\oval(0.1,0.1)[br]}    % 0 -0.03
\put(4,2.42){\oval(0.1,0.1)[tl]}       % -0.04 -0.12
\put(4.74,3.03){\oval(0.1,0.1)[tl]}     % startNE
\put(4.79,3.15){\oval(0.1,0.1)[br]}    % +0.05 +0.12
\put(4.89,3.18){\oval(0.1,0.1)[tl]}      % +0.1 +0.03
\put(4.94,3.3){\oval(0.1,0.1)[br]}      % +0.05 +0.12
\put(4.74,3.0){\oval(0.1,0.1)[bl]}      % startSE
\put(4.79,2.88){\oval(0.1,0.1)[tr]}     % +0.05 -0.12
\put(4.89,2.85){\oval(0.1,0.1)[bl]}     % +0.1 -0.03
\put(4.94,2.73){\oval(0.1,0.1)[tr]}     % +0.05 -0.12
\put(5.5,2.91){$+\q g^3$}
\put(6.9,3){\vector(-1,0){0.1}}
\multiput(6.9,3)(0.175,0){8}{\line(1,0){0.0875}}
\put(8.3,3){\line(1,0){1.4}}
\put(7.7,2.99){\oval(0.1,0.1)[bl]}      % startS
\put(7.74,2.87){\oval(0.1,0.1)[tr]}     % +0.04 -0.12
\put(7.74,2.84){\oval(0.1,0.1)[br]}    % 0 -0.03
\put(7.7,2.72){\oval(0.1,0.1)[tl]}        % -0.04 -0.12
\put(7.7,2.69){\oval(0.1,0.1)[bl]}       % 0 -0.03
\put(7.74,2.57){\oval(0.1,0.1)[tr]}     % +0.04 -0.12
\put(7.74,2.54){\oval(0.1,0.1)[br]}    % 0 -0.03
\put(7.7,2.42){\oval(0.1,0.1)[tl]}       % -0.04 -0.12
\put(8.4,2.99){\oval(0.1,0.1)[bl]}      % startS
\put(8.44,2.87){\oval(0.1,0.1)[tr]}     % +0.04 -0.12
\put(8.44,2.84){\oval(0.1,0.1)[br]}    % 0 -0.03
\put(8.4,2.72){\oval(0.1,0.1)[tl]}        % -0.04 -0.12
\put(8.4,2.69){\oval(0.1,0.1)[bl]}       % 0 -0.03
\put(8.44,2.57){\oval(0.1,0.1)[tr]}     % +0.04 -0.12
\put(8.44,2.54){\oval(0.1,0.1)[br]}    % 0 -0.03
\put(8.4,2.42){\oval(0.1,0.1)[tl]}       % -0.04 -0.12
\put(9.1,2.99){\oval(0.1,0.1)[bl]}      % startS
\put(9.14,2.87){\oval(0.1,0.1)[tr]}     % +0.04 -0.12
\put(9.14,2.84){\oval(0.1,0.1)[br]}    % 0 -0.03
\put(9.1,2.72){\oval(0.1,0.1)[tl]}        % -0.04 -0.12
\put(9.1,2.69){\oval(0.1,0.1)[bl]}       % 0 -0.03
\put(9.14,2.57){\oval(0.1,0.1)[tr]}     % +0.04 -0.12
\put(9.14,2.54){\oval(0.1,0.1)[br]}    % 0 -0.03
\put(9.1,2.42){\oval(0.1,0.1)[tl]}       % -0.04 -0.12
\put(9.74,3.03){\oval(0.1,0.1)[tl]}     % startNE
\put(9.79,3.15){\oval(0.1,0.1)[br]}    % +0.05 +0.12
\put(9.89,3.18){\oval(0.1,0.1)[tl]}      % +0.1 +0.03
\put(9.94,3.3){\oval(0.1,0.1)[br]}      % +0.05 +0.12
\put(9.74,3.0){\oval(0.1,0.1)[bl]}      % startSE
\put(9.79,2.88){\oval(0.1,0.1)[tr]}     % +0.05 -0.12
\put(9.89,2.85){\oval(0.1,0.1)[bl]}     % +0.1 -0.03
\put(9.94,2.73){\oval(0.1,0.1)[tr]}     % +0.05 -0.12
\put(10.5,2.91){$+\q g^3$}
\put(11.9,3){\vector(-1,0){0.1}}
\multiput(11.9,3)(0.175,0){12}{\line(1,0){0.0875}}
\put(14,3){\line(1,0){0.76}}
\put(12.7,2.99){\oval(0.1,0.1)[bl]}      % startS
\put(12.74,2.87){\oval(0.1,0.1)[tr]}     % +0.04 -0.12
\put(12.74,2.84){\oval(0.1,0.1)[br]}    % 0 -0.03
\put(12.7,2.72){\oval(0.1,0.1)[tl]}        % -0.04 -0.12
\put(12.7,2.69){\oval(0.1,0.1)[bl]}       % 0 -0.03
\put(12.74,2.57){\oval(0.1,0.1)[tr]}     % +0.04 -0.12
\put(12.74,2.54){\oval(0.1,0.1)[br]}    % 0 -0.03
\put(12.7,2.42){\oval(0.1,0.1)[tl]}       % -0.04 -0.12
\put(13.4,2.99){\oval(0.1,0.1)[bl]}      % startS
\put(13.44,2.87){\oval(0.1,0.1)[tr]}     % +0.04 -0.12
\put(13.44,2.84){\oval(0.1,0.1)[br]}    % 0 -0.03
\put(13.4,2.72){\oval(0.1,0.1)[tl]}        % -0.04 -0.12
\put(13.4,2.69){\oval(0.1,0.1)[bl]}       % 0 -0.03
\put(13.44,2.57){\oval(0.1,0.1)[tr]}     % +0.04 -0.12
\put(13.44,2.54){\oval(0.1,0.1)[br]}    % 0 -0.03
\put(13.4,2.42){\oval(0.1,0.1)[tl]}       % -0.04 -0.12
\put(14.1,2.99){\oval(0.1,0.1)[bl]}      % startS
\put(14.14,2.87){\oval(0.1,0.1)[tr]}     % +0.04 -0.12
\put(14.14,2.84){\oval(0.1,0.1)[br]}    % 0 -0.03
\put(14.1,2.72){\oval(0.1,0.1)[tl]}        % -0.04 -0.12
\put(14.1,2.69){\oval(0.1,0.1)[bl]}       % 0 -0.03
\put(14.14,2.57){\oval(0.1,0.1)[tr]}     % +0.04 -0.12
\put(14.14,2.54){\oval(0.1,0.1)[br]}    % 0 -0.03
\put(14.1,2.42){\oval(0.1,0.1)[tl]}       % -0.04 -0.12
\put(14.84,3.03){\oval(0.1,0.1)[tl]}     % startNE
\put(14.89,3.15){\oval(0.1,0.1)[br]}    % +0.05 +0.12
\put(14.99,3.18){\oval(0.1,0.1)[tl]}      % +0.1 +0.03
\put(15.04,3.3){\oval(0.1,0.1)[br]}      % +0.05 +0.12
\put(14.84,3.0){\oval(0.1,0.1)[bl]}      % startSE
\put(14.89,2.88){\oval(0.1,0.1)[tr]}     % +0.05 -0.12
\put(14.99,2.85){\oval(0.1,0.1)[bl]}     % +0.1 -0.03
\put(15.04,2.73){\oval(0.1,0.1)[tr]}     % +0.05 -0.12
\put(0.5,0.91){$+ \q \cO(g^4)\ .$}
\end{picture}
\\
The Feynman-like rules for these graphs are as follows:

\setlength{\unitlength}{1cm}
\thicklines
\begin{picture}(16,6)
%\put(-1,0){\line(0,1){6}}
%\put(-1,0){\line(1,0){16}}
%\put(-1,6){\line(1,0){16}}
%\put(15,0){\line(0,1){6}}
%
\put(1,5){\vector(-1,0){1}}
\put(2,4.91){$\overleftarrow{\frac{\delta}{\delta A_\mu}}\,\gamma_{\mu\nu}\,\partial^\nu C$}
\put(7.5,5){\vector(-1,0){0.1}}
\multiput(7.5,5)(0.2,0){8}{\line(1,0){0.1}}
\put(8.4,4.99){\oval(0.1,0.1)[bl]}      % startS
\put(8.44,4.87){\oval(0.1,0.1)[tr]}     % +0.04 -0.12
\put(8.44,4.84){\oval(0.1,0.1)[br]}    % 0 -0.03
\put(8.4,4.72){\oval(0.1,0.1)[tl]}        % -0.04 -0.12
\put(8.4,4.69){\oval(0.1,0.1)[bl]}       % 0 -0.03
\put(8.44,4.57){\oval(0.1,0.1)[tr]}     % +0.04 -0.12
\put(8.44,4.54){\oval(0.1,0.1)[br]}    % 0 -0.03
\put(8.4,4.42){\oval(0.1,0.1)[tl]}       % -0.04 -0.12
\put(10.1,4.91){$-2\,\overleftarrow{\frac{\delta}{\delta A_\mu}}\,\delta_{\mu\nu}^{\alpha\beta}\,\partial_\alpha C\,\partial^\nu(A_\beta\,C)$}
\put(1,3){\line(-1,0){1}}
\put(1.8,2.91){$S_0\,=\,-\,\slashed{\partial}\,C$}
\multiput(5.8,3)(0.2,0){5}{\line(1,0){0.1}}
\put(7.5,2.91){$G_0\,=\,-\,C$}
\put(12,3){\line(-1,0){0.5}}
\put(12.04,3.03){\oval(0.1,0.1)[tl]}     % startNE
\put(12.09,3.15){\oval(0.1,0.1)[br]}    % +0.05 +0.12
\put(12.19,3.18){\oval(0.1,0.1)[tl]}      % +0.1 +0.03
\put(12.24,3.3){\oval(0.1,0.1)[br]}      % +0.05 +0.12
\put(12.04,3.0){\oval(0.1,0.1)[bl]}      % startSE
\put(12.09,2.88){\oval(0.1,0.1)[tr]}     % +0.05 -0.12
\put(12.19,2.85){\oval(0.1,0.1)[bl]}     % +0.1 -0.03
\put(12.24,2.73){\oval(0.1,0.1)[tr]}     % +0.05 -0.12
%\put(13,2.91){$\sfrac12\,\slashed{A}^2$}
\put(12.9,2.91){$\sfrac12\,\gamma^{\rho\lambda}\,A_\rho\,A_\lambda$}
\put(1.2,1.3){\line(-1,0){1.2}}
\put(0.6,1.29){\oval(0.1,0.1)[bl]}      % startS
\put(0.64,1.17){\oval(0.1,0.1)[tr]}     % +0.04 -0.12
\put(0.64,1.14){\oval(0.1,0.1)[br]}    % 0 -0.03
\put(0.6,1.02){\oval(0.1,0.1)[tl]}        % -0.04 -0.12
\put(0.6,0.99){\oval(0.1,0.1)[bl]}       % 0 -0.03
\put(0.64,0.87){\oval(0.1,0.1)[tr]}     % +0.04 -0.12
\put(0.64,0.84){\oval(0.1,0.1)[br]}    % 0 -0.03
\put(0.6,0.72){\oval(0.1,0.1)[tl]}       % -0.04 -0.12
\put(1.8,0.91){$-\,\slashed{A}$}
\put(6.9,1.3){\line(-1,0){0.6}}
\multiput(5.8,1.3)(0.2,0){3}{\line(1,0){0.1}}
\put(6.4,1.29){\oval(0.1,0.1)[bl]}      % startS
\put(6.44,1.17){\oval(0.1,0.1)[tr]}     % +0.04 -0.12
\put(6.44,1.14){\oval(0.1,0.1)[br]}    % 0 -0.03
\put(6.4,1.02){\oval(0.1,0.1)[tl]}        % -0.04 -0.12
\put(6.4,0.99){\oval(0.1,0.1)[bl]}       % 0 -0.03
\put(6.44,0.87){\oval(0.1,0.1)[tr]}     % +0.04 -0.12
\put(6.44,0.84){\oval(0.1,0.1)[br]}    % 0 -0.03
\put(6.4,0.72){\oval(0.1,0.1)[tl]}       % -0.04 -0.12
\put(7.6,0.91){$-\,\slashed{A}$}
\multiput(11.6,1.3)(0.2,0){6}{\line(1,0){0.1}}
\put(12.2,1.29){\oval(0.1,0.1)[bl]}      % startS
\put(12.24,1.17){\oval(0.1,0.1)[tr]}     % +0.04 -0.12
\put(12.24,1.14){\oval(0.1,0.1)[br]}    % 0 -0.03
\put(12.2,1.02){\oval(0.1,0.1)[tl]}        % -0.04 -0.12
\put(12.2,0.99){\oval(0.1,0.1)[bl]}       % 0 -0.03
\put(12.24,0.87){\oval(0.1,0.1)[tr]}     % +0.04 -0.12
\put(12.24,0.84){\oval(0.1,0.1)[br]}    % 0 -0.03
\put(12.2,0.72){\oval(0.1,0.1)[tl]}       % -0.04 -0.12
\put(13.4,0.91){$-\,\pa^\nu A_\nu$}
\end{picture}
Furthermore, a trace has to be performed in spinor space.
Since for each fermion line, gamma matrices from the free fermion propagators~$S_0$
and from the photon insertions~$\slashed{A}$ are multiplied along the linear tree, 
the trace short-circuits the tree in spinor space and contracts the Lorentz indices of
the partial derivatives on~$C$ and of the photon emissions~$A$ in every possible fashion.

\vskip 0.3cm
\noindent 
The perturbative map 
\begin{align}
(\cT_g\,A)_\mu^a(x) \ =\ A_\mu^a(x)\ +\ \sum_{n=1}^\infty \frac{g^n}{n!}\,(T_n\,A)_\mu^a(x)
\end{align}
is obtained by
iterating the $\mathcal{R}$ operation to build $(\cT^{-1}_g A)_\mu^a(x)$
according to \eqref{eq:TInverse} and inverting the power series (see~\eqref{eq:inversion}). 
In terms of the power-series components $R_n$ defined above, 
the $n$-th order contribution to $\cT_g \,A$ is given by~\cite{L2}
\begin{align}
T_n\,A \ =\ \sum_{\{n\}} c_{\{n\}} \,R_{n_s} R_{n_{s-1}} \cdots R_{n_2}  R_{n_1}\,A\, ,
\end{align}
with the sum running over all multiindices
\begin{align}
\{n\} = \bigl( n_1, n_2, \ldots, n_s )  \qq \text{with} \q \sum_{i=1}^s n_i = n \q \text{and} \q n_i\in\natural
\end{align}
and with coefficients
\begin{align}
c_{\{n\}} \ =\ (-1)^s\frac{n!}{n_1\cdot (n_1{+}n_2) \cdot (n_1{+}n_2{+}n_3) \cdots (n_1{+}\cdots{+}n_s)}\, .
\end{align}
To order $g^4$, this expansion reads
\begin{align}
\begin{aligned}
\cT_g\,A \ &=\ A \ -\ g\,R_1 A \ -\ \sfrac12g^2\bigl(R_2-R_1^2\bigr)A\ -\ 
\sfrac16g^3\bigl(2R_3-R_1R_2-2R_2R_1+R_1^3\bigr)A \\
&\q -\sfrac{1}{24}g^4\bigl(6R_4-2R_1R_3-3R_2R_2+R_1^2R_2-6R_3R_1
+2R_1R_2R_1+3R_2R_1^2-R_1^4\bigr)A \\
&\q +\ {\cal O}(g^5)\, .
\end{aligned}
\end{align}
The repeated action of $\mathcal{R}$ on itself grafts linear trees onto linear trees.
This produces binary trees of all kinds with double leaves $\slashed{A}^2$,
where multiple gamma-matrix traces run over all possible parts of those trees, 
and any part of a tree may have fermion lines replaced by ghost lines.
Excluded only are length-one ghost lines and ghost lines ending in a double leaf.
After performing the gamma-matrix traces, all lines become scalar propagators $C$ dressed
with a partial derivative, whose Lorentz indices get contracted in almost all possible ways.
However, there appear partial cancellations of gamma-matrix traces between trees of the same topology.
The combinatorial factors and Feynman rules for these trees will be given elsewhere.

\vskip 0.3cm
\noindent 
Finally, one may raise the question of the uniqueness of $\cT_g A$. 
As the map is constructed iteratively from the $\mathcal{R}$~operator, 
a non-uniqueness will originate from an ambiguity in~$\mathcal{R}$. 
In $D{=}4$, such an ambiguity arises from the freedom to add a topological term $\sim\int F{\wedge}\,F$
to the bosonic action, which allows for the modification
\begin{align}
\slashed{A}^2 \q \longmapsto \q \slashed{A}^2\,\bigl(1+\kappa\,\c^5\bigr)
\end{align}
in the $\mathcal{R}$ operator and, hence, in the Feynman rules. 
As a consequence, novel terms carrying the epsilon tensor appear in the expansion of~$\cT_gA$.
This offers, for $\kappa=\pm1$, the option of a chiral Nicolai map~\cite{L1,Nic2}.
The possibility of such a chiral projection may be explored for the other critical dimensions as well.

\newpage
\appendix
\section{Proof of the main theorem} 
\label{sec:FlumeLechtenfeldProof}
In this appendix we firstly construct the $\mathcal{R}$ operator from the response of the vacuum expectation value 
of an arbitrary product of bosonic operators to changes in the coupling constant. 
This generalizes the argument first presented in \cite{FL} for the $D{=}4,\,\cN{=}1$
theory to all other pure supersymmetric Yang--Mills theories.
Secondly, we prove the distributivity of the $\mathcal{R}$ operation. 
Thirdly, we recall the argument that $\mathcal{R}$  annihilates the gauge-invariant bosonic Yang--Mills action 
as well as the gauge-fixing function.  
These properties imply the determinant matching and suffice to establish the main theorem.

\subsection{Construction of the \texorpdfstring{$\cR$}{R} operator} 
In this section we show how to construct the $\mathcal{R}$ operator for any pure 
super Yang--Mills theory. The action consists of a gauge-invariant part \cite{SYM1}
\begin{align}\label{Sinv}
S_\text{inv}\ =\ \sfrac{1}{4} \int \md x \ F_{\m \n}^a(x) F_{\m \n}^a(x) 
\ +\ \sfrac{1}{2} \int \md x \ \bar{\l}^a(x)\,\c^\m (D_\m \l)^a(x) 
\end{align}
and a gauge-fixing part
\begin{align}\label{Sgf}
S_\text{gf}\ =\  \frac{1}{2\xi} \int \md x \  G^a(A)\,G^a(A)
\ +\ \int \md x \  \bar{C}^a(x)\, \frac{\pa G^a(A)}{\pa A_\mu^b(x)}  (D_\m C)^b(x) \, .
\end{align}
The full  action $S_\text{inv} + S_\text{gf}$ is invariant under the BRST (or Slavnov) variations
\begin{align}\label{s}
s A_\m^a = (D_\m C)^a\,, \q
s \l^a = -g f^{abc} \l^b C^c\,, \q
s C^a = -\sfrac{g}2 f^{abc} C^b C^c\,,\q
s \bar{C}^a = -\sfrac1{\x}\, G^a(A)
\end{align}
for all positive $\xi$ and an arbitrary gauge-fixing function $G^a(A)$ (which for simplicity we assume not to depend on $g$).
In the remainder we will specialize to the Landau gauge function (\ref{Landau}), {\em i.e.} $G^a=\pa^\mu\! A_\mu^a$. 
This is the so-called $R_\xi$ gauge; the Landau gauge is obtained for $\xi\ra 0$. 
For the ghost kinetic term we thus recover the standard form
\begin{equation}
\int \md x \  \bar{C}^a(x)\, \frac{\pa G^a(A)}{\pa A_\mu^b(x)}  (D_\m C)^b(x)
\ = \  \int \md x \  \bar{C}^a(x)\,\pa^\m (D_\m C)^a(x) \, .
\end{equation}

%\vskip 0.3cm
\noindent
For an arbitrary product $X$ of bosonic operators the linear response of its vacuum expectation value to a change in the coupling constant is given by
\begin{align}
\frac{\md}{\md g} \bigl< X \bigr> \ =\ 
\frac{\md}{\md g} \bigl<\!\!\bigl< X \bigr>\!\!\bigr> \ =\ 
\biggl<\!\!\!\!\biggl<  \frac{\md X}{\md g} \biggr>\!\!\!\!\biggr> 
- \biggl<\!\!\!\!\biggl< \frac{\md ( S_\text{inv}+ S_\text{gf} ) }{\md g}\ X \biggr>\!\!\!\!\biggr>
\ =:\ \bigl< \mathcal{R}\ X \bigr> \, .
\end{align}
Here, the vacuum expectation values 
$\big\langle\!\!\big\langle \cdots \big\rangle\!\!\big\rangle$ 
and $\big\langle\cdots\big\rangle$ were defined in (\ref{corr0}), and we have dropped 
subscripts $g$ or $\xi$ for simplicity of notation.
Making use of supersymmetry, 
we want to rewrite the right-hand side in terms of a derivational operator~$\mathcal{R}$.
To this end we introduce
\begin{align}
\D_\a \ =\ - \frac{1}{2r} f^{abc} \int \md x \ \left( \c^{\r \l} \l^a(x) \right)_\a A_\r^b(x) A_\l^c(x) 
\end{align}
and use the standard supersymmetry variations (with supersymmetry 
parameter stripped off)
\begin{align}
\d_\a \l^a_\b \ =\ \sfrac{1}{2} ( \c^{\m \n} )_{\b \a} F^a_{\m \n} \qq\text{and}\qq
\d_\a A^a_\m \ =\ - ( \bar{\l}^a \c_\m )_\a
\end{align}
to compute
\begin{align}
\d_\a \D_\a \ =\ \sfrac12 f^{abc} \int \md x\ F_{\m\n}^a(x) A_\m^b(x) A_\n^c(c) 
\ +\ \sfrac{D-1}{r} f^{abc} \int \md x\ \bigl( \c^{\m } \l^a(x)  \bar{\l}^b(x)  \bigr)_{\a \a}  A_\m^c(x)
\end{align}
so that
\begin{align}
\frac{\md S_\text{inv}}{\md g}\ =\ \d_\a \D_\a 
\ +\ \bigl(\sfrac12-\sfrac{D-1}{r}\bigr) f^{abc} \int \md x\ \bar{\l}^a(x) \c^\m \l^b(x) A_\m^c(x)\, .
\end{align}
Notice that $\d_\a$ anticommutes with other anticommuting operators. With
\begin{align}
\frac{\md S_\text{gf}}{\md g}\ =\ f^{abc} \int \md x \ \bar{C}^a(x)\,\pa^\m \bigl( A_\m^b(x)\,C^c(x) \bigr)
\end{align}
we arrive at
\begin{align}
\begin{aligned}
\frac{\md}{\md g} \left< X \right>
\ &=\ \biggl<\!\!\!\!\biggl< \frac{\md X}{\md g} \biggl>\!\!\!\!\biggl> 
\ -\ \Bigl<\!\!\!\Bigl< \left( \d_\a \D_\a \right) X \Bigl>\!\!\!\Bigl>  \\
&\q +\ \biggl<\!\!\!\!\biggl< \left(\sfrac{D-1}{r} - \sfrac12 \right)  f^{abc}
\int \md x \  \bar{\l}^a(x) \c^\m A_\m^c(x) \l^b(x) \ X \biggl>\!\!\!\!\biggl>  \\
&\q -\ \biggl<\!\!\!\!\biggl< f^{abc} \int \md x \ \bar{C}^a(x)\, \pa^\m
\big( A_\m^b(x)\, C^c(x) \big) \ X \biggl>\!\!\!\!\biggl> \, .
\label{eq:ContructROpIntermed1}
\end{aligned}
\end{align}

\vskip 0.3cm
\noindent 
We want to rewrite
\begin{align}
 \bigl<\!\!\bigl< ( \d_\a \D_\a ) X \bigl>\!\!\bigl>\ =\  
 \bigl<\!\!\bigl< \D_\a \d_\a X \bigr>\!\!\bigl>\  +\  \bigl<\!\!\bigl< \d_\a ( \D_\a X )  \bigr>\!\!\bigl>
\label{eq:ContructROpIntermed2}
\end{align}
using the supersymmetry Ward identity
\begin{align}
\bigl<\!\!\bigl<  \d_\a Y \bigr>\!\!\bigl> \ =\ \bigl<\!\!\bigl< (\d_\a S_\text{gf} )\, Y \bigr>\!\!\bigl> \, .
\end{align}
Employing the Slavnov variations (\ref{s}) one finds that
\begin{align}
\d_\a S_\text{gf} \ =\ - s \int \md x \ \bar{C}^a(x) \, \d_\a \bigl( \pa^\m A_\m^a(x) \bigr) \, .
\end{align}
Thus, the Ward identity becomes
\begin{align}
\bigl<\!\!\bigl< \d_\a Y \bigr>\!\!\bigl> \ =\   
\biggl<\!\!\!\!\biggl< \int \md x \ \bar{C}^a(x) \, \d_\a \bigl( \pa^\m A_\m^a (x) \bigr) \ s( Y) \biggl>\!\!\!\!\biggl> \, .
\end{align}
By \eqref{eq:ContructROpIntermed2} we need to apply this to $Y=\D_\a X$, and from 
$s(\D_\a X) = s(\D_\a)X-\D_\a s(X)$ we also require the Slavnov variation of $\D_\a$. 
Making use of the Jacobi identity we get
\begin{align}
s \left( \D_\a \right) \ =\   \frac{1}{r} f^{abc} \int \md x \ \bigl( \c^{\r \l} \l^a(x) \bigr)_\a  \pa_\r C^b(x) A_\l^c(x) \, .
\end{align}
Subsequently we can put everything back together,
\begin{align}
\frac{\md}{\md g} \bigl< X \bigr> \ =\  
\biggl< \frac{\md X}{\md g} \biggl>\ -\ 
\bigl<\!\!\bigl< \D_\a \d_\a X \bigl>\!\!\bigl> \ +\ 
\biggl<\!\!\!\!\biggl<  \int \md x \ \bar{C}^a(x)\,\d_\a \bigl( \pa^\m A_\m^a(x) \bigr)  \D_\a \, s( X) \biggl>\!\!\!\!\biggl>
\ +\ \bigl<\!\!\bigl< Z\ X\bigr>\!\!\bigr>
\end{align}
with
\begin{align}
\begin{aligned}
Z\ =\ &-\int \md y \  \bar{C}^a(y)\,\d_\a \bigl( \pa^\m A_\m^a(y) \bigr) \
\frac{1}{r} f^{bcd}  \int \md x \ \bigl( \c^{\r \l } \l^b(x) \bigr)_\a A_\r^c(x)\,\pa_\l C^d(x)  \\
&+\ \left(\sfrac{D-1}{r} - \sfrac12 \right) f^{abc} \int \md x  \  \bar{\l}^a(x) \c^\m A_\m^c(x) \l^b(x)
\ -\ f^{abc} \int \md x \  \bar{C}^a(x)\, \pa^\m \big( A_\m^b(x)\,C^c(x)\big)\, .
\end{aligned}
\label{eq:Z}
\end{align}
As it stands, and up to this point, the above derivation 
is valid {\em for all values of the gauge parameter\/} $\xi$. We can therefore take the limit
$\xi\ra 0$, for which all contributions containing $\pa^\mu\!A_\mu^a$ simply vanish (recall
that physical quantities anyway cannot depend on $\xi$). 
We will show in the next subsection that under these conditions the multiplicative contribution disappears,
\begin{align}
\lim_{\xi\ra 0} \; \bigl<\!\!\bigl< Z\ X\bigr>\!\!\bigr>_\xi 
= 0 \qq \text{for} \q \frac{D-1}{r} - \frac12 =\frac1r \, ,
\label{eq:FlumeLechtenfeldClaim}
\end{align}
and thus only in the critical dimensions $D=3,4,6$ and $10$, where $r = 2(D{-}2)$ indeed.
Therefore, by integrating over all fermionic degrees of freedom we finally obtain
\begin{align}
\mathcal{R} \, X \ =\ \frac{\md X}{\md g}\ +\  \bcontraction{}{\d}{_\a X \cdot}{\D} \d_\a X \cdot \D_\a\ +\ \int \md x \ \bcontraction[2ex]{}{\bar{C}}{^a(x)\ \bcontraction{}{\d}{_\a \bigl( \pa^\m A_\m^a(x) \bigr) \, }{\D} \d_\a \bigl( \pa^\m A_\m^a(x) \bigr) \, \D_\a \, }{s} \bar{C}^a (x)\ \bcontraction{}{\d}{_\a \bigl( \pa^\m A_\m^a(x) \bigr) \, }{\D} \d_\a \bigl( \pa^\m A_\m^a(x) \bigr) \, \D_\a \, s(X) \, ,
\label{eq:ROperatorVersion1}
\end{align}
where the contractions signify fermionic (gaugino or ghost) propagators in the gauge-field background.
For $\mathcal{N}=1$ super Yang--Mills theory this result was first derived in 
\cite{FL}, see also \cite{Nic2}.

\subsection{Distributivity of the \texorpdfstring{$\cR$}{R} operation}
In this subsection we generalize the argument from \cite{FL}  in order to prove that
\eqref{eq:FlumeLechtenfeldClaim} holds for any bosonic functional~$X$,
subject to the conditions stated above, allowing us
to ignore all terms containing $\pa^\mu\!A_\mu^c$.
Integrating out the gauginos and ghosts, this amounts to the relation
\begin{align}
\bigl< \, \bcontraction{}{}{\!\!Z}{}Z\ X \bigr> = 0 \qq\forall\,X 
\qq\Leftrightarrow\qq \bcontraction{}{}{\!\!Z}{}Z \,=\, 0\, .
\end{align}
Functionally integrating \eqref{eq:Z} over the gaugino and ghost degrees of freedom we derive
\begin{align}
\begin{aligned}
\bcontraction{}{}{\!\!Z}{}Z &\ =\ \frac{1}{r} \int \md y \ \bigl(  \bcontraction[3ex]{}{\bar{C}}{^a(y) \bigr) \bigl( \pa_\r^y \bar{\l}^a(y) \c^\r \bigr)_\a \ f^{bcd} \int \md x \bigl( \c^{\m \n} \l^c(x) \bigr)_\a A_\m^c(x)\,\pa_\n^x }{C} \bar{C}^a(y) \bigl) \bigr(\pa_\r^y \bcontraction[2ex]{}{\bar{\l}}{^a(y) \c^\r \bigr)_\a \  f^{bcd} \int \md x \  \  \bigl( \c^{\m \n} }{\l} \bar{\l}^a (y)\c^\r \bigr)_\a \  f^{bcd} \int \md x \ \bigl( \c^{\m \n} \l^b(x) \bigr)_\a A_\m^c(x)\,\pa_\n^x C^d(x) \\
&\q +\ \left(\sfrac{D-1}{r} - \sfrac12 \right)  f^{abc} \int \md x \  \bcontraction{}{\bar{\l}}{_\a^a(x) \c_{\a \b}^\m   A_\m^b(x) }{\l} \bar{\l}_\a^a(x) \c_{\a \b}^\m A_\m^b(x) \l_\b^c(x) \ -\ f^{abc} \int \md x \ \bcontraction{}{\bar{C}}{^a(x) A_\m^b (x) \pa_x^\m }{C} \bar{C}^a(x) A_\m^b (x) \pa_x^\m C^c(x) \, .
\end{aligned}
\end{align}
We use the identity
$\c^{\m \n} = \frac{1}{2} \left(  \c^\m \c^\n -  \c^\n \c^\m \right) =  - \c^\n \c^\m  + \d^{\m \n}$ 
and reorder the contracted terms so as to identify any contraction with a fermion or ghost propagator 
(in the presence of the gauge-field background) to get
\begin{align}
\begin{aligned}
\bcontraction{}{}{\!\!Z}{}Z &\ =\   -\frac{1}{r} f^{bcd} \int \md x \ \md y \  \tr \left(  \pa_\n^x G^{da}(x,y) \c^\r  \c^\n  \c^{\m}  \pa_\r^y S^{ba}(x,y) \right) A_\m^c(x) \\
&\qq +\ \frac{1}{r} f^{bcd} \int \md x \ \md y \  \tr \left( \pa_\n^x G^{da}(x,y)   \c^\r \d^{\m \n} \pa_\r^y S^{ba}(x,y) \right) A_\m^c(x) \\
&\qq -\ \left(\sfrac{D-1}{r} - \sfrac12 \right) f^{abc} \int \md x \ \tr \left( S^{ca}(x,x) \c^\m  \right) A_\m^b(x) 
\ +\ f^{abc} \int \md x \ \pa_x^\m G^{ca}(x,x) A_\m^b(x)   \, .
\end{aligned}
\end{align}
Then we need the following Schwinger--Dyson identities, which follow directly from \eqref{eq:fullprops} 
and the relation $\c^\n\pa_\n^x G_0^{da}(x{-}y)=S_0^{da}(x{-}y)$, namely
\begin{align}
\begin{aligned}
S^{ba}(x,y) &\ =\ S_0^{ba}(x{-}y)\ +\ g f^{emn}  \int \md z \  S_0^{be}(x{-}z)  A_\n^n(z) \c^\n S^{ma}(z,y) \, ,\\
\c^\n \pa_\n^x G^{da}(x,y)  &\ =\  S_0^{da}(x{-}y)\ +\ g f^{emn} \int \md z \  S_0^{de}(x{-}z)  A_\n^n(z) \pa_z^\n G^{ma}(z,y)  \, .
\end{aligned}
\end{align}
Integrating by parts and using $\c^\r \pa_\r^y S_0^{da}(x{-}y) = -  \d^{da} \d(x{-}y)$ 
together with $\tr \  {\bf{1}} = r$, this gives
\begin{align}
\begin{aligned}
\bcontraction{}{}{\!\!Z}{}Z &\ =\   - \frac{1}{r} f^{bca} \int \md x \  \tr \left(   \c^\m S^{ba}(x,x) \right) A_\m^c(x) \\
&\q -\  \frac{g}{r} f^{bcd} f^{emn} \int \md x \ \md y \ \md z \   \tr \left( S_0^{de}(x{-}z) A_\n^n(z) \pa_z^\n G^{ma}(z,y)   \c^{\m}  \pa_\r^y S^{ba}(x,y) \c^\r \right) A_\m^c(x) \\
&\q -\  f^{acd} \int \md x  \  \pa_x^\m G^{da}(x,x)   A_\m^c(x) \\
&\q +\ \frac{g}{r} f^{bcd}  f^{emn}\int \md x \ \md y \ \md z \  \tr \left( \pa_x^\m G^{da}(x,y)  S_0^{be}(x{-}z) \c^\n A_\n^n(z)  \pa_\r^y S^{ma}(z,y)  \c^\r \right) A_\m^c(x) \\
&\q -\ \left(\sfrac{D-1}{r} - \sfrac12 \right) f^{abc} \int \md x \ \tr \left( S^{ca}(x,x) \c^\m  \right) A_\m^b(x) 
\ +\ f^{abc} \int \md x \ \pa_x^\m G^{ca}(x,x) A_\m^b(x) \, 
\end{aligned}
\end{align}
(the formally singular terms with coincident arguments can be appropriately regulated, if needed). The pure fermion loops (first and fifth term) cancel, provided (\ref{rD}) holds with
$D=3,4,6$ or $10$, as advertised. The pure ghost loops (third and sixth term) 
cancel independently of dimension. 
Finally, we use $S_0^{be}(x{-}z) =- S_0^{be}(z{-}x)$ to cancel the two remaining terms,
\begin{align}
\begin{aligned}
\bcontraction{}{}{\!\!Z}{}Z   &\ =\   -\frac{g}{r} f^{bcd} f^{emn} \int \md x \ \md y \ \md z \   \tr \left( S_0^{de}(x{-}z) A_\n^n(z) \pa_z^\n G^{ma}(z,y)   \c^{\m}  \pa_\r^y S^{ba}(x,y) \c^\r \right) A_\m^c(x) \\
&\q +\ \frac{g}{r} f^{bcd}  f^{emn}\int \md x \ \md y \ \md z \  \tr \left( \pa_x^\m G^{da}(x,y)  S_0^{be}(x{-}z) \c^\n A_\n^n(z)  \pa_\r^y S^{ma}(z,y)  \c^\r \right) A_\m^c(x) \\
&\q =\ 0 \, .
\end{aligned}
\end{align}
This concludes the proof.

\subsection{\texorpdfstring{$\cR$}{R} annihilates the bosonic action and the gauge-fixing function}

Statement $[3.]$ in the main theorem, about determinant matching, follows from the other two parts and from 
the equality~\eqref{corr1} relating interacting and free-field correlators.
The latter can be seen to be equivalent~\cite{L2}  to a fixed-point property of $\cT_g$ under the coupling constant flow,
\begin{align}
\mathcal{R}\ \cT_g A\ =\ 0\, ,
\end{align}
from which in fact $T_g$ can be constructed directly~\cite{L2}. 

\vskip 0.3cm
\noindent
Statement [1.] in the main theorem, $S_g[A]=S_0[\cT_gA]$, is equivalent to a property of the kernel of~$\mathcal{R}$, namely,
\begin{align}
\mathcal{R}\ S_g[A]\ =\ 0\, .
\end{align}
For completeness, let us recall the proof that $\mathcal{R}$ annihilates the bosonic 
invariant action \cite{Nic2}.
Recall the form \eqref{eq:ROperatorVersion2} of the $\mathcal{R}$ operator 
with the covariant transversal projector~\eqref{eq:covproj}. From
\begin{align}
\frac{\md S_g}{\md g}\ =\ \frac12 f^{abc} \int \md x\ F_{\m\n}^a(x) A_\m^b(x) A_\n^c(x)
\qq\text{and}\qq
\frac{\d\,S_g[A]}{\d A_\m^a(x)}\ =\ -\bigl( D_\s F^{\s\m}\bigr)^a(x)
\end{align}
we obtain
\begin{align}
\begin{aligned}
\mathcal{R}\,S_g[A]\ &=\
\frac12 f^{abc} \int \md x\ F_{\m\n}^a(x) A_\m^b(x) A_\n^c(x) \\
&\, +\ \frac{1}{2r} \int \md x\,\md z\, \md y \  \bigl( D_\s F^{\s\m}\bigr)^a(x) \,P_{\m\n}^{ae}(x,z)\,
\tr \left( \c^\n S^{eb}(z,y)\,\c^{\r \l} \right)\,f^{bcd} A_\r^c(y) A_\l^d(y) \, .
\end{aligned}
\label{eq:RS}
\end{align}
We can take advantage of the fact that $S_g$ is gauge invariant: Since
\begin{align}
\int\md x\ \bigl( D_\s F^{\s\m}\bigr)^a(x) \,\bigl(D_\m G\bigr)^{ae}(x,z)\,\pa_\n
\ =\  -\int\md x\ ( D_\m D_\s F^{\s\m}\bigr)^a(x) \,G^{ae}(x,z)\,\pa_\n \ =\ 0\, ,
\end{align}
the projector $P_{\m\n}$ in $\mathcal{R}$ can be replaced by the identity. 
Then the second term in \eqref{eq:RS} becomes
\begin{align}
\frac{1}{2r} \int \md x\ \md y \  \bigl( D^\s F_{\s\n}\bigr)^a(x)\
\tr \left( \c^\n S^{ab}(x,y)\,\c^{\r \l} \right)\,f^{bcd} A_\r^c(y) A_\l^d(y)\, .
\label{eq:term2}
\end{align}
To bring the first term in a similar form, we use the identity
\begin{align}
\d_{\m\n}^{\r\l}\,\d^{ab}\,\d(x{-}y)\ =\ -\frac{1}{2r} \tr \left(\c_{\m\n} \c^\s (D_\s S)^{ab}(x,y)\,\c^{\r\l}\right)
\end{align}
to blow it up to
\begin{align}
\begin{aligned}
&\frac12 \int \md x\ F_{\m\n}^a(x)\,\d_{\m\n}^{\r\l}\,\d^{ab}\,f^{bcd} A_\m^c(x) A_\n^d(x) \\
&\qq = -\frac{1}{4r}\int \md x\ \md y \  F_{\m\n}^a(x)\ \tr\left(\c^{\m\n} \c^\s (D_\s S)^{ab}(x,y)\,\c^{\r\l}\right) 
f^{bcd}A_\m^c(y) A_\n^d(y) \\
&\qq = \ \ \frac{1}{4r}\int \md x\ \md y \  (D_\s F_{\m\n})^a(x)\ \tr\left(\c^{\m\n} \c^\s S^{ab}(x,y)\,\c^{\r\l}\right) 
f^{bcd}A_\m^c(y) A_\n^d(y) \\
&\qq = -\frac{1}{4r}\int \md x\ \md y \  (D_\s F_{\m\n})^a(x) \
\tr\left(\bigl[2\c^\n \d^{\m\s}+\c^{\n\m\s}\bigr] S^{ab}(x,y)\,\c^{\r\l}\right) f^{bcd}A_\m^c(y) A_\n^d(y) \\
&\qq = -\frac{1}{2r} \int \md x\ \md y \  \bigl( D^\s F_{\s\n}\bigr)^a(x)\
\tr \left( \c^\n S^{ab}(x,y)\,\c^{\r \l} \right)\,f^{bcd} A_\r^c(y) A_\l^d(y)\, ,
\end{aligned}
\label{eq:term1}
\end{align}
where we partially integrated $D_\s$, employed 
$\c^{\m\n}\c^\s=2\c^{[\m}\d^{\n]\s}+\c^{\m\n\s}$ and observed the Bianchi identity
$\c^{\s\m\n}D_\s F_{\m\n}=0$.
In this form, the first term \eqref{eq:term1} is seen to cancel the second term~\eqref{eq:term2}, 
which proves the assertion.

\vskip 0.3cm
\noindent 
Statement [2.] in the main theorem is almost trivial, given the form \eqref{eq:covproj} of the
transversal projector inside~$\mathcal{R}$.
The fixed-point feature of the gauge-fixing function, 
\begin{align}
G^a(T_g A)\ =\ G^a(A) \qq\Leftrightarrow\qq \mathcal{R}\ G^a(A)\ =\ 0\, ,
\end{align}
is built into the formalism, since the projector by construction annihilates~$G^a$.
For the Landau gauge, 
\begin{align}
\frac{\pa G^a(A)}{\pa A_\m^b(x)}\,P^{bc}_{\m\n}(x,z) 
\ =\ \pa^\m_x \Bigl( \d^{ac}\d_{\m\n}\d(x{-}z) - (D_\m G)^{ac}(x,z) \pa_\nu^z \Bigr)\ =\ 0
\end{align}
by the definition of the ghost propagator.
This is also apparent from the form~\eqref{eq:ROperatorVersion1} of the $\mathcal{R}$ operator,
where for $X=G^a(A)$ the second and third terms cancel each other. 
In fact, this consideration generalizes to an arbitrary gauge.\footnote{
We tacitly assume that $\frac{\md G}{\md g}=0$. For $g$-dependent gauges one must reconsider.}

\vskip 0.3cm
\noindent 
The main theorem is herewith proved, for the Landau gauge.

%\newpage
\section{Explicit expression for \texorpdfstring{$\cR^3$}{R3}} \label{sec:R3}
In this section we give the explicit expression for $\mathcal{R}^3\bigl( A_\m^a(x) \bigr)$. Applying the $\mathcal{R}$-operator three times and repeatedly using \eqref{eq:ROperatorVersion2} as well as \eqref{eq:SgDerivative} and \eqref{eq:SVariation} we arrive at
\begin{align}
\begin{aligned}
&\mathcal{R}^3\bigl( A_\m^a(x) \bigr) \Big\vert_{g=0} \\
&= \bigg\{  - \frac{1}{4r^3} \, \tr \left(\c^\m \c^{\c \tau} \c^\n \right) \tr \left( \c^\c \c^\x \c^\b \c^\a \right) \tr \left( \c^\tau \c^\l \c^\s \c^\r \right) \\
& \qqq   + \frac{1}{2r^2} \, \tr \left(\c^\m \c^\x \c^\b \c^{\a} \c^\c  \c^\n \right) \tr \left(\c^\c \c^\s \c^\l \c^\r \right) \bigg\} \  f^{abc} f^{bde} f^{cmn} \\
& \qq \times \int \md y \ \md z \ \md w \ \pa_\n C(x-y) \pa_\r C(y-z) A_\l^d(z) A_\s^e(z) \pa_\a C(y-w) A_\b^m(w) A_\x^n(w) \\
&\q - \frac{1}{2r^2} \, \tr \left(\c^\x  \c^\l \c^\r \c^\n \right) \tr \left( \c^\b \c^\a \c^{\s} \c^\x \right) f^{abc} f^{bde} f^{cmn} \\
& \qq \times \int \md y \ \md z \ \md w \ \pa_\m C(x-y) \pa_\n C(y-z) A_\r^d(z) A_\l^e(z) \pa_\s C(y-w) A_\a^m(w) A_\b^n(w) \\ 
&\q + \bigg\{ \frac{1}{2r^3} \, \tr \left(\c^\m \c^{\r \c} \c^\n \right) \tr \left( \c^\c \c^{\s \tau} \c^\l \right) \tr \left( \c^\tau \c^\x \c^\b \c^\a \right) \\
&\qqq  - \frac{1}{2r^2} \, \tr \left(\c^\m \c^{\l \r} \c^\n \right) \tr \left( \c^\x \c^\b \c^\a\c^\s \right) \\
&\qqq - \frac{1}{r^2} \, \tr \left(\c^\m \c^{\s \c} \c^\l  \c^\r \c^\n \right) \tr \left( \c^\c\c^\x \c^\b \c^\a \right) \\
&\qqq   - \frac{1}{2r^2} \, \tr \left( \c^\m \c^{\r \c} \c^\n \right) \tr \left(\c^\c \c^\x \c^\b \c^\a  \c^\s \c^\l \right) \\
&\qqq  + \frac{1}{r}\, \tr \left( \c^\m \c^\x \c^\b \c^\a \c^\s \c^\l \c^\r \c^\n \right)  \bigg\} \  f^{abc} f^{bde} f^{dmn} \\
&\qq  \times \int \md y \ \md z \ \md w \ \pa_\n C(x-y) A_\r^d(y) \pa_\l C(y-z) A_\s^e(z) \pa_\a C(z-w) A_\b^m(w) A_\x^n(w) \\
&\q + \bigg\{  \frac{1}{r^2} \, \tr \left(\c^\x \c^\b \c^\a \c^\s \right) \tr \left( \c^{\l \x} \c^\r  \c^\n \right)  - \frac{1}{r} \, \tr \left( \c^\n \c^\b \c^\a \c^\s \c^\l \c^\r \right) \bigg\} \  f^{abc} f^{bde} f^{dmn} \\
&\qq \times \int \md y \ \md z \ \md w \ \pa_\m C(x-y) A_\n^d(y) \pa_\r C(y-z) A_\l^e(z) \pa_\s C(z-w) A_\a^m(w) A_\b^n(w) \\
&\q  +  \frac{1}{r}   \tr \left(  \c^\r  \c^\l  \c^\a \c^{\s}    \right) \ f^{abc}  f^{bde} f^{dmn} \\
&\qq \times   \int \md y \ \md z \  \md w \   \Pi_{\m \n} (x-z) A_\n^c(z) C(z-y)  A_\r^e(y) \pa_\l C(y-w)   A_\s^m(w) A_\a^n(w) \, .
\label{eq:R3}
\end{aligned}
\end{align}
To evaluate the $\gamma$-traces 
$t^{\mu_1\cdots \mu_{2n}} \equiv {\rm Tr}\, \gamma^{\mu_1} \cdots \gamma^{\mu_{2n}}$
we use (\ref{id}) and the standard recursion relation
\begin{equation}
t^{\mu_1\cdots \mu_{2n+2}} \ =\ \delta^{\mu_1\mu_2} t^{\mu_3\mu_4\cdots \mu_{2n+2}}
       \  -\  \delta^{\mu_1\mu_3} t^{\mu_2\mu_4\cdots \mu_{2n+2}} \ \pm \  \ldots
\end{equation}
(with $2n{+}1$ terms on the r.h.s., so the full trace has altogether $(2n{+}1)!!$ terms). 
Possible ambiguities related to topological terms mentioned at the end of section~4 
and related to chiral spinors and additional $\ve$-tensors in these traces will be discussed elsewhere.

\vskip 0.3cm
\noindent
Computing the traces and collecting all terms produces a total of 45 terms,
\begin{align}
\begin{aligned}
&\mathcal{R}^3\bigl( A_\m^a(x) \bigr) \Big\vert_{g=0} \\
&=-2   f^{a b c} f^{b d e} f^{c m n} \int \md y \ \md z \  \md w  \ \pa_\r C(x-y) \  \bigg\{ \\
&\hspace{2cm} + \pa_\l C(y-z) A_\l^d(z) A_\s^e(z)  \pa_\m C(y-w) A_\r^m(w) A_\s^n(w) \\ 
&\hspace{2cm} + \pa_\l C(y-z)  A_\m^d(z) A_\r^e(z) \pa_\s C(y-w) A_\l^m(w) A_\s^n(w)   \\ 
&\hspace{2cm} +3 \, \pa_\l C(y-z)  A_\m^d(z) A_\l^e(z) \pa_\s C(y-w) A_\r^m(w) A_\s^n(w)   \\ 
&\hspace{2cm}+ \pa_\r C(y-z)  A_\m^d(z) A_\s^e(z) \pa_\l C(y-w) A_\l^m(w) A_\s^n(w)  \smash{\bigg\}} \\ 
&\q - 6  f^{a b c} f^{b d e} f^{d m n}  \int \md y \ \md z \  \md w  \ \pa_\r C(x-y)A_\l^c(y) \ \bigg\{ \\
&\hspace{2cm} + 2 \,   \pa_{[\r} C(y-z) A_{\s]}^e(z) \pa_{[\l} C(z-w) A_\m^m(w) A_{\s]}^n(w)  \\ 
&\hspace{2cm} - 2 \, \pa_{[\l} C(y-z) A_{\s]}^e(z)   \pa_{[\r} C(z-w) A_\m^m(w) A_{\s]}^n(w) \\ 
&\hspace{2cm}-   \pa_\s C(y-z) A_\s^e(z)\pa_{[\r} C(z-w)  A_\m^m(w) A_{\l]}^n(w)  \\ 
&\hspace{2cm} - 2 \,  \pa_{[\s} C(y-z) A_{\m]}^e(z) \pa_{[\r} C(z-w) A_\l^m(w) A_{\s]}^n(w)  \\ 
&\hspace{2cm} +  4 \, \pa_{[\r} C(y-z) A_\m^e(z)  \pa_{|\s|} C(z-w) A_{\l]}^m(w) A_\s^n(w)  \smash{\bigg\}}  \\  
&\q + 3  f^{a b c} f^{b d e} f^{d m n}  \int \md y \ \md z \  \md w  \ \pa_\r C(x-y) \  \bigg\{ \\
&\hspace{2cm} + 3 A_{\m}^c(y)  \pa_{\l} C(y-z) A_{\s}^e(z)  \pa_{[\r} C(z-w)  A_\l^m(w) A_{\s]}^n(w)  \\  
&\hspace{2cm} +  3 A_{\r}^c(y)  \pa_{\l} C(y-z) A_{\s}^e(z)  \pa_{[\l} C(z-w)  A_\m^m(w) A_{\s]}^n(w) \\
&\hspace{2cm}-  4A_{[\m}^c(y) \pa_{|\s}C(y-z) A_\s^e(z) \pa_\l C(z-w) A_{\l|}^m(w) A_{\r]}^n(w)   \\ 
&\hspace{2cm}-  4 A_{[\m}^c(y) \pa_{|\l|} C(y-z) A_{\r]}^e(z)  \pa_\s C(z-w)  A_\l^m(w) A_\s^n(w)  \smash{\bigg\}} \\ 
&\q - 2  f^{a b c} f^{b d e} f^{d m n}  \int \md y \ \md z \  \md w  \   \bigg\{ \\
&\hspace{2cm} + 2 \pa_\r C(x-y) A_{[\r}^c(y) \pa_{\m]} C(y-z) A_\l^e(z)  \pa_\s C(z-w)  A_\l^m(w) A_\s^n(w)   \\ 
&\hspace{2cm} -  \pa_\m C(x-y) \,  \pa_\r  \left( A_\r^c(y) C(y-w) \right) A_\l^m(z) A_\s^n(z)  \pa_\l C(z-w) A_\s^e(w)  \smash{\bigg\}}  \\ 
&\q + 2  f^{a b c} f^{b d e} f^{d m n}  \int \md y \ \md z \  
A_\m^c(x) C(x-y)  A_\r^e(y) \pa_\l C(y-z) A_\r^m(z) A_\l^n(z)  \ .
\end{aligned}
\end{align}
It is only after inverting the full series that we obtain the somewhat more compact result \eqref{eq:MainResult}. 
We also note that,  before computing the traces, $T_3 A$ has new and additional terms
as compared to $\mathcal{R}^3 A$. Only the factors in \eqref{eq:R3} change so that 
upon taking the trace more terms cancel.

\vskip 0.3cm
\noindent 
Finally we remark that the above expansion for $\cT_g^{-1}A$ is the one needed for the computation of 
quantum correlators at $\cO(g^3)$, in order to extend the recent rederivation of certain 
$\cN=4$ correlators up to $\cO(g^2)$ in \cite{NP}.

\vskip 0.3cm
\noindent
{\bf Acknowledgements:} We thank the referees for their critical comments on a previous 
version  and for their insistence on further clarification of several important points.

\vskip 1cm

%\newpage

\end{document}